\documentclass[npg]{copernicus}

\newcommand{\aap}{Astron. Astrophys.}
\newcommand{\apj}{Astrophys. J.}
\newcommand{\apjl}{Astrophys. J. Lett.}

\newcommand{\araa}{Annu. Rev. Astron. Astrophys.}

\newcommand{\mnras}{Mon. Not. R. Astron. Soc.}
\newcommand{\nat}{Nature}
\newcommand{\prl}{Phys. Rev. Lett.}
\newcommand{\jgr}{J. Geophys. Res.}

\newcommand{\planss}{Plan. Space Sci.}
\newcommand{\pra}{Phys. Rev. A}

\begin{document}

\title{Reconnection in Turbulent Fluids and Reconnection Diffusion: Implications for Star Formation}
\author{A. Lazarian}
\affil{Department of Astronomy, University of Wisconsin, 475 North Charter Street, Madison, WI 53706, USA}

\runningtitle{Reconnection and Star Formation}
\runningauthor{A. Lazarian}

\correspondence{A. Lazarian (lazarian@astro.wisc.edu)}

\maketitle

\begin{abstract}
Astrophysical fluids are turbulent a fact which changes the dynamics of many key processes, including magnetic
reconnection. Fast reconnection of magnetic field in turbulent fluids allows the field to change its topology and 
connections. As a result, the traditional concept of magnetic fields being frozen into the plasma is no longer
applicable. Plasma associated with a given magnetic field line at one instant is distributed along a different set of magnetic field lines
at the next instant. This diffusion of plasmas and magnetic field is enabled by reconnection and therefore is termed "reconnection diffusion". 
The astrophysical implications of this concept include heat transfer in plasmas, advection of heavy elements in interstellar medium, magnetic
field generation etc. However, the most dramatic implications of the concept are related to the star formation process. The reason is that magnetic
fields are dynamically important for most of the stages of star formation.  The existing theory
of star formation has been developed ignoring the possibility of reconnection diffusion. Instead, it appeals to the decoupling of mass and magnetic field arising from neutrals drifting in 
respect to ions entrained on magnetic field lines, i.e. through the process 
that is termed "ambipolar diffusion". The predictions of ambipolar diffusion and reconnection diffusion are very
different. For instance, if the ionization of media is high, ambipolar diffusion predicts that the coupling of mass and magnetic field is nearly perfect. At the same time, reconnection diffusion 
is independent of the ionization but depends on the scale of the 
turbulent eddies and on the turbulent velocities. 
In the paper we explain the physics of reconnection diffusion both from macroscopic and microscopic points of view, i.e. appealing to the reconnection of flux tubes and
to the diffusion of magnetic field lines. We quantify the reconnection diffusion rate both for weak and strong MHD turbulence and address the problem of reconnection diffusion
acting together with ambipolar diffusion. In addition, we provide a criterion for correctly representing the magnetic diffusivity in simulations
of star formation.  We discuss 
the intimate relation between the processes of reconnection diffusion, field wandering and turbulent mixing of a magnetized media and
show that the role of the plasma effects is limited to "breaking up lines" on small scales and does not affect the
rate of reconnection diffusion. We address the existing observational results and demonstrate how reconnection
diffusion can explain the puzzles presented by observations, in particular, the observed higher magnetization of cloud cores
in comparison with the magnetization of envelopes. We also outline a possible set of observational tests of the
reconnection diffusion concept and discuss how the application of the new concept changes our understanding of 
star formation and its numerical modeling.  Finally, we outline the differences of the process 
of reconnection diffusion and the process of accumulation of matter along magnetic field lines that is frequently
invoked to explain the results of numerical simulations. 
\end{abstract}

\introduction  

Magnetic flux freezing is a key textbook concept with a huge impact on astrophysical theory. The concept was 
first proposed by Alfven (1942) whose principle of Òfrozen-inÓ field lines has provided a powerful heuristic 
influencing our understanding of many astrophysical processes such as star formation, stellar collapse, evolution of accretion disks, magnetic dynamo etc. This principle, however, is not universal, as we discuss below, its violation entails important consequences for
star formation.

The phenomenon of fast magnetic reconnection is an example of failure of the magnetic freezing concept. Fast magnetic
reconnection is a type of reconnection which does not depend on resistivity and thus should proceed in a media of negligible 
resistivity. Solar flares provide an example of the rapid 
energy release which would not be  possible if magnetic fields were perfectly frozen in (see Yamada, Kulsrud \& Ji 2010 and references 
therein).

The issue of "flux freezing" violation by magnetic reconnection has been known for some time, but was not taken very seriously due to the unclear nature of fast reconnection (see Zweibel \& Yamada 2009 and references therein). Indeed, for years it was considered that fast reconnection required some special physical conditions and therefore the "flux freezing " is fulfilled everywhere apart from some special zones.

The present day star formation paradigm has been developed that flux freezing concept holds. In magnetically-mediated star formation theory which was founded by the pioneering studies by L. Mestel  and L. Spitzer (see Mestel \& Spitzer 1956, Mestel 1966) and  brought to the level of sophistication by other researchers (see Shu, Adams \& Lizano 1987, Mouschovias 1991, Nakano et al. 2002, Shu et al. 2004, Mouschovias et al. 2006). According to
it, magnetic fields slow down and even prevent star formation if the media is sufficiently magnetized. In the 
{\it assumption} of flux freezing of the magnetic field in the ionized component, the change of the flux to mass ratio happens due
to neutrals which do not feel magnetic field directly, but only through ion-neutral interactions. In the presence of gravity, neutrals get concentrated towards the center of the gravitational potential while magnetic fields resist compression and therefore leave the
forming protostar (e.g. Mestel 1965). This makes star formation inefficient for magnetically dominated (i.e. subcritical) clouds. The low efficiency of star formation corresponds to observations (e.g. Zuckerman \& Evans 1974), which is usually interpreted as a strong argument in support of the above scenario. This does not solve all the problems as, at the same time, for clouds dominated by gravity, i.e. supercritical clouds, this scenario does not work as magnetic fields do not have time to leave the cloud through ambipolar diffusion. Therefore for 
supercritical clouds magnetic field should be dragged into the star, forming stars with magnetizations far in excess of the observed ones (see Galli et al. 2006, Johns-Krull 2007).   

What we described above is the initial stage of star formation. However, magnetic fields are important for other stages of star formation, e.g. formation of the accretion disks around forming stars. This, as we discuss below, is also problematic if one relies
on ambipolar diffusion. In fact, classical ideas of star formation based exclusively on ambipolar diffusion have been challenged by observations (Troland \& Heiles 1986, Shu et al. 2006, Crutcher et al. 2009, 2010, see Crutcher 2012 for a review).
While the interpretation of particular observations is the subject of scientific debates (see Moschovias \& Tassis 2009), it is suggestive that there may be additional processes that the classical theory does not take into account. The primary suspect is turbulence, which is ubiquitous in the interstellar media and molecular clouds (see Larson 1981, Armstrong et al. 1994, Elmegreen \& Falgarone 1996, Lazarian \& Pogosyan 2000, Stanimirovic \& Lazarian 2001, Heyer \& Brunt 2004, Padoan et al. 2006, 2009, Chepurnov \& Lazarian 2010, Burkhart et al. 2010). Turbulence has revolutionized the field of star formation (see Vazquez-Semadeni et al. 1995, Ballesteros-Paredes et al. 1999, Elmegreen 2000, 2002, McKee \& Tan 2003, Elmegreen \& Scalo 2004, MacLow \& Klessen 2004, McKee \& Ostriker 2007) but the treatment of the turbulent magnetic fields stayed within the flux freezing paradigm.

The understanding of flux freezing in turbulent astrophysical environments has been challenged relatively recently and not
all the consequences of this radical change have been evaluated so far. Lazarian \& Vishniac (1999, henceforth LV99) identified magnetic field wandering, which is inherent property of magnetized turbulent plasma, as the cause of fast, i.e. independent of resistivity, magnetic reconnection. They showed that in turbulent fluids magnetic fields should undergo constant reconnection and change their identity all the time. This implies that magnetic fields are not any more frozen into a perfectly conducting fluid if this fluid is turbulent  as was explicitly stated first in Vishniac \& Lazarain (1999). Later, the challenge to the concept of "flux freezing" came from another side, i.e. from more formal mathematical studies of 
magnetic fields properties in turbulent fluids (see Eyink 2011a).  Eyink, Lazarian \& Vishniac (2011, henceforth ELV11) showed the consistency of these two approaches and established the equivalence of the LV99 treatment with that in more recent mathematical papers.  

While the idea that turbulence can change the reconnection rates has been discussed in a number of earlier papers, the LV99 model was radically different from its predecessors. For instance, 
Mathaeus \& Lamkin (1985, 1986) performed 2D numerical
simulations of turbulence and provided arguments in favor of magnetic reconnection getting fast. However, the physics of the
processes that they considered was very different from that in LV99 (see more \S 5.3).

The LV99 analytical predictions have been successfully tested in Kowal et al. (2009) which made it important to study astrophysical applications of the model. Magnetic reconnection was treated in LV99 for both collisional and collisionless turbulent plasmas and was extended to the partially ionized gas in Lazarian, Vishniac \& Cho (2004). This motivated Lazarian (2005) to identify fast reconnection of magnetic field as an essential process of removing magnetic flux at different stages of star formation. Later, in Lazarian et al. (2010) this process was termed "reconnection diffusion". 

One may claim that the reconnection diffusion concept extends the concept of hydrodynamic turbulent diffusion to magnetized fluids. The physics of it is very different from the "magnetic turbulent
diffusivity" idea discussed within the theories of kinematic dynamo (see Parker 1979). Reconnection diffusion, unlike "magnetic turbulent diffusivity", deals with dynamically important
magnetic fields, e.g. with subAlfvenic turbulence. Thus magnetic fields are {\it not} passively mixed and magnetic reconnection plays a vital role for the process. Recently, the consequences of
reconnection diffusion have been studied numerically for the diffuse interstellar media, molecular clouds and accretion disks (Santos-Lima et al. 2010, 2012). 
These studies should be understood in the appropriate context. For example, it is wrong to view numerical calculations in Santos-Lima et al. (2010, 2012) as the 
actual justification of the reconnection diffusion concept.  The
concept of reconnection diffusion can only be justified appealing to the LV99 model and to high resolution numerical testing of the latter. Indeed, the
testing of the reconnection rates predicted in LV99 is performed at much higher resolution (see Kowal et al. 2009, 2012, Lazarian
et al. 2011) than the resolution of the cores and accretion disks in our simulations in Santos-Lima et al. (2010, 2012). In other words, the numerical
effects are under control in the simulations focused to test LV99 theoretical predictions, while the LV99 theory justifies why one should not be 
too much worried about small scale numerical effects present in the astrophysically-movitated set ups exploring astrophysical consequences of reconnection diffusion. 

The major shortcoming of our previous papers on reconnection diffusion is that the physical picture of the process
has never been clearly explained.  This causes confusion and prevents the heuristic use of the reconnection diffusion concept.
The situation is aggravated by the fact that the textbook picture of reconnection involves magnetic fluxes of 
opposite polarities getting into contact. This is clearly different from what one expects to see during the star
formation process where magnetic fields without large scale reversals are being dragged together towards the center
of the gravitational potential.  

The goal of this paper is to clarify what are the actual foundations of the reconnection diffusion concept,
present the physical picture of the diffusion of magnetic field and plasmas in the presence of turbulence,
to provide the comparison of the predictions based on the reconnection diffusion concept and observations
as well as to present testable predictions. We attempt to provide foundations for the alternative picture of the star formation process 
based on the process of reconnection diffusion.

In what follows, we discuss the problem of magnetic flux removal for star formation in \S 2, present the description
of MHD turbulence in \S 3, describe the model of magnetic reconnection in turbulent media in \S 4, introduce the concept
of reconnection diffusion in \S 5. In \S 6 we discuss the microscopic picture of reconnection diffusion and in \S 7 consider a possibility of two distinct regimes of reconnection diffusion. \S 8 outlines the limitations of numerical studies and presents the numerical results of reconnection diffusion for different stages of star formation; in \S 9 we discuss how the existing observational data corresponds to the reconnection diffusion model and provide observational predictions. Additional
consequences of reconnection diffusion for observations and numerics are outlined in \S 10, while the discussion and the conclusions
are presented in \S 11 and \S 12, respectively.  


\section{Star Formation and Magnetic Flux Problem}

\subsection{Turbulence in magnetized interstellar plasmas}

A paradigm shift, a concept popularized by the science historian Thomas Kuhn (1962), usually delivers a major advance in our perception of reality. In their 2004 ARA\&A review, Scalo \& Elmegreen write: ''One of the most important developments in the field of interstellar gas dynamics during the last half century was renewed perception that most processes and structures are strongly affected by turbulence. This is a paradigm shift unparalleled in many other fields of astronomy, comparable perhaps to the discovery of extrasolar planets and cosmological structure at high redshift.''

It has been known for decades that interstellar medium (ISM) is driven by violent supernovae explosions (McKee \& Ostriker 1977). By now it has been accepted that the ISM is turbulent on scales ranging from AUs to kpc (see Armstrong et al. 1995, Elmegreen \& Scalo 2004, Lazarian 2009). Fig. 1 shows the turbulent power density plotted against the inverse
of the scale length, with data at large scales, i.e. at small wavenumbers $q$ expanded using the Wisconsin H$_{\alpha}$ Mapper (WHAM) data on electron density fluctuations (Chepurnov \& Lazarian 2010). 

A more direct evidence comes from the observations of spectral lines. Apart from showing non-thermal Doppler broadening (see Larson 1981), they also reveal spectra of supersonic turbulent velocity fluctuations when analyzed with techniques like Velocity Channel Analysis (VCA) of Velocity Coordinate Spectrum (VCS) developed (see Lazarian \& Pogosyan 2000, 2004, 2006, 2008) and applied to the observational data (see Padoan et al. 2004, 2009, Chepurnov et al. 2010).

\begin{figure}
\centering
  \includegraphics[height=.45\textheight]{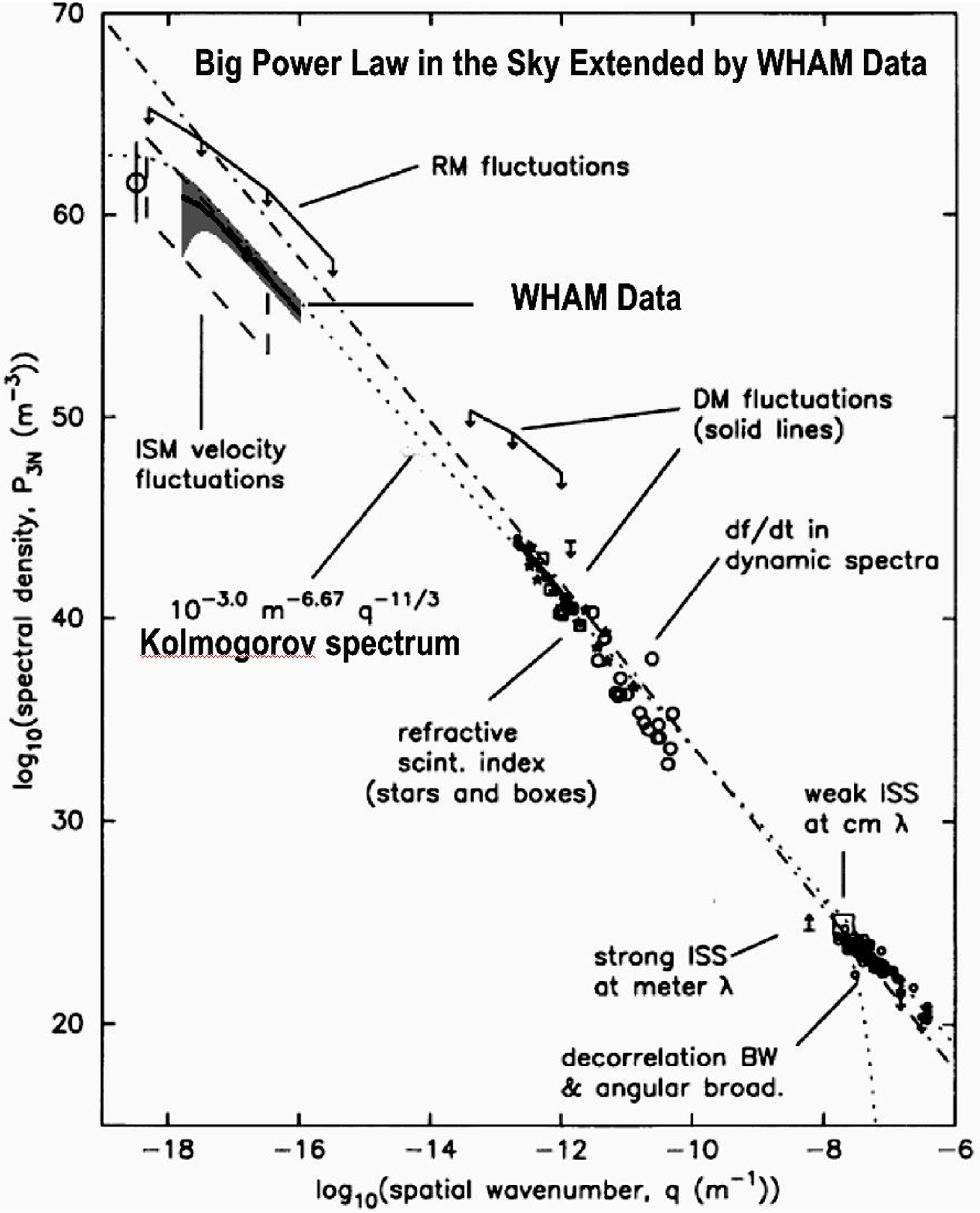}
  \caption{Turbulence in the interstellar gas as revealed by electron density fluctuations. "Big Power Law in the Sky" in Armstrong et al. (1995) extended using WHAM data. The slope corresponds to that
of Kolmogorov turbulence. Modified from Chepurnov \& Lazarian (2010).}
 \label{CL}
\end{figure}

We should clarify that turbulence in astrophysical environments is the most natural and expected phenomenon.  Magnetized astrophysical plasmas generally have very large Reynolds numbers due to the large length scales involved, as well as the fact that the motions of charged particles in the direction perpendicular to magnetic fields are constrained. Laminar plasma flows at these high Reynolds numbers $R=VL_f/\nu$, where $V$ and $L_f$ are the velocity
and the scale of the flow, $\nu$ is fluid viscosity,  are prey to numerous linear and finite-amplitude instabilities, from which turbulent motions readily develop.
The drivers of turbulence include supernovae explosions that shape the interstellar medium (McKee \& Ostriker 1977, Nakamura et al. 2006),
accretion flows (see Klessen \& Hennebelle 2010), magneto-rotational instability in the galactic disk (Sellwood \& Balbus 1999), thermal instability (see Kritsuk \& Norman 2002, Koyama \& Inutsuka 2002), collimated outflows (see Nakamura \& Li 2007) etc. In fact, present day understanding of interstellar medium views molecular clouds as part of interstellar turbulent cascade, i.e. as density fluctuations in a compressible flow (see Ostriker,
Stone, \& Gammie 2001, McKee \& Ostriker 2007 and references therein). As molecular clouds are magnetized, one must understand the diffusion of magnetic fields in turbulent flows in order to describe star formation. 

\subsection{Star formation and role of turbulence}

Star formation is known to be a rather inefficient process. Indeed, the mass of the Milky way is $M_{MW}\approx10^9$ solar mass. For the typical density of the gas of 50 cm$^{-3}$ the free fall time is $\tau_{ff}\approx
(3\pi/32G\rho)^{1/2}\approx 6\times 10^6$ years which provides a "natural" star formation rate $M_{MW}/\tau_{ff}$ of 200 solar mass per year. At the same time the measured star formation rate is $\approx$ 1.3 solar
mass per year (Murray \& Rahman 2010). It is because of this inefficiency we still have large amounts of present in interstellar media of spiral galaxies. A traditional way of explaining this inefficiency is to appeal to 
magnetic forces preventing gravitational collapse (see Mouschovias \& Spitzer 1976).  

The role of magnetic fields has been subjected to scrutiny from the very beginning of research on star formation theory (see Mestel \& Spitzer 1956). Eventually, star formation happens due to gravity, but if the portion of gas is strongly magnetized, the magnetic
field may prevent such a collapse. It is possible to show that if the ratio of the magnetic flux to mass is larger than the  critical one:
\begin{equation}
(\Phi/M)_{crit}\approx 1.8\times 10^{-3} ~{\rm gauss}~{\rm cm}^2~g^{-1}
\label{crit}
\end{equation}
magnetic field prevents the cloud collapse (see Draine 2011). These strongly magnetized clouds are termed {\it subcritical}. Such clouds cannot collapse unless than lose their magnetic flux. The typical values of magnetization
of diffuse gas corresponds to subcritical magnetization. Therefore the compression of matter together with magnetic field would result in subcritical clouds. Naturally, if most
of the clouds are subcritical, star formation is inefficient. 

To address the problem of the magnetic field diffusion both in the partially ionized ISM and in molecular clouds, researchers usually appeal to the ambipolar diffusion concept (see Mestel 1956, Shu 1983). The idea of the ambipolar diffusion is very simple and may be easily exemplified in the case of gas collapsing to form a protostar. As the magnetic field is acting on charged particles only, it does not directly affect neutrals. Neutrals move under the gravitational pull but are scattered by collisions with ions and charged dust grains that are coupled with the magnetic field. The resulting flow dominated by the neutrals does not drag the magnetic field lines and those will diffuse away dragging through the infalling matter. This process of ambipolar diffusion becomes faster as the ionization ratio decreases and therefore, becomes more important in poorly ionized cloud cores. The corresponding theory of star
formation based on ambipolar diffusion is well developed (see Mouschovias et al. 2006) and it predicts low star formation rates
in agreement with observations. 

If the cloud is not magnetically dominated, i.e. it is supercritical, the above discussion is not applicable and the gravity should
induce a collapse dragging magnetic flux with entrained matter into the forming star on the timescales less than the ambipolar
diffusion time. This presents twofold problem. First of all, simple estimates show that if all the magnetic flux is brought together with the material that collapses to form a star in molecular clouds, then the magnetic field in a proto-star should be several orders of magnitude higher than the one observed in stars (this is the ``magnetic flux problem'', see Galli et al. (2006) and references therein). 
For instance, T-Tauri stars have magnetic field $\approx 2\times 10^3$ Gauss (see Johns-Krull 2007), which amounts to $(\Phi/M)\approx 3\times 10^{-8}$ gauss~cm$^2$~g$^{-1}$, which is a million times smaller
that the flux to mass ratio estimated for one solar mass clump in a cloud of density $10^4$ cm$^{-3}$ (see Draine 2011). To avoid the "magnetic flux problem" one has to identify ways of efficient magnetic flux removal.  

As we discuss earlier, turbulence is an essential process for interstellar media.
Traditionally, in the textbooks the role of turbulence was mostly limited to affecting the virial mass.  Numerical simulations have shown that turbulence can play the dominant role
for the formation of the molecular clouds (Ballestros-Paredes et al. 2007). Moreover, simulations were indicating the ability of turbulence to change the flux to mass ratio even without ambipolar diffusion.
 Those were interpreted as suggestive of a scenario in which
compressible turbulence collects matter along magnetic field lines and induces supercritical star formation (Vazquez-Semadeni et al. 2011).
One might claim that this approach does not require magnetic flux diffusion provided that
 the collapse is strictly one-dimensional. As we discuss in \S 11.6 this scenario is very restrictive and
in this paper we propose a different solution of the magnetic flux problem for star formation, namely, we claim that the turbulent scenario should
be extended by allowing the process of reconnection diffusion that is inevitably induced by MHD turbulence.

There have been attempts to "turbocharge" ambipolar diffusion by combining its action with turbulence. We briefly discuss 
these ideas in \S 11.4. Our claim in the paper that in the presence of turbulence the removal of magnetic flux is fast and
ambipolar diffusion is not required.

\subsection{Problems of the ambipolar diffusion paradigm}

The alternatives to the classical star formation theory have emerged in the last decade as numerical simulations showed
 that star formation can also be slow due to feedback introduced by turbulence (see McKee \& Ostriker 2007, Pudritz \& Kevlahan 2012). This feedback can disperse clouds before they can give birth to new stars. The fact that clouds and cores must be necessarily subcritical has also been challenged both numerically (see Padoan et al. 2004) and observationally (see Crutcher 2012 and references therein). In fact, in his 2012 ARA\&A review Crutcher states: "There is no definitive evidence for subcritical molecular clouds or for ambipolar diffusion driven star formation." Below we list some other examples where researchers claim that ambipolar diffusion is not adequate to explain observations. 

Shu et al. (2006) explored the accretion phase in low-mass star formation and concluded that ambipolar diffusion could work only under  rather special circumstances like, for instance, considering particular dust grain sizes. Instead they proposed a solution which is based on the magnetic flux diffusion by Ohmic resistivity. However,
to do the job, they postulated that the Ohmic resistivity should be increased by about 4 orders of magnitude. We feel that there is no physical justification for such a resistivity enhancement (see more in \S 11.5).  

Observations of different regions of the diffuse ISM compiled by Troland \& Heiles (1986) indicate that magnetic fields and density are not correlated. Such a correlation would be expected in a naive picture of turbulence where compressions
of magnetic fields are accompanied by compressions of density. Magnetic diffusion can explain the absence of correlations, but the ambiplar diffusion in diffuse ISM is negligible due to high degree of gas ionization. In addition, observations of magnetized cores by Crutcher et al. (2010) contradict to the predictions of the ambipolar diffusion paradigm. All these cases are troublesome 
from the point of ambipolar diffusion paradigm but, as we discuss further in the paper
(see \S 9), are consistent with the reconnection diffusion concept that we advocate.

\section{Turbulent astrophysical media and its description}

To quantify reconnection diffusion the quantitative description of astrophysical turbulence is required. 

\subsection{Magnetized turbulence in astrophysical plasmas}

In addition to being turbulent, astrophysical plasmas are magnetized (see Spitzer 1978, Draine 2011). The magnetization of astrophysical fluids most frequently arises from the dynamo action to which turbulence is an essential component (see Schekochihin et al. 2007). In fact, it has been shown that turbulence converts in weakly magnetized conducting fluid from five to ten percent of the energy of the cascade into the magnetic field energy (see Cho et al. 2009). 
This fraction {\it does not} depend on the original magnetization\footnote{This makes the problem of the initial or seed magnetic field, that for a long time has worried researchers, rather trivial. Very weak magnetic fields, e.g. generated by Bierman battery
(see Lazarian 1992) can be amplified fast in a turbulent plasmas.} and therefore magnetic fields will come to equipartition\footnote{In supersonic
flows compressibility effects induce deviations from the equipartition.} with the turbulent motions within a few eddy turnover times.

We deal with magnetohydrodynamic (MHD) turbulence which provides a correct fluid-type description of plasma turbulence at large scales (see \S 4.2). Astrophysical turbulence is a direct consequence of large scale fluid motions experiencing low friction. The Reynolds numbers are typically very large in astrophysical flows as the scales are large. As magnetic fields decrease the viscosity for the plasma motion perpendicular to their direction, $Re$ numbers get really astronomically large. For instance, $Re$ numbers of $10^{10}$ are very common for astrophysical flow. For so large $Re$ the inner degrees of fluid motion get excited and a complex pattern of motion develops.

The drivers of turbulence, e.g. supernovae explosions in the interstellar medium, inject energy at large scales and then the energy cascades down to small scales through the hierarchy of eddies spanning up over the entire inertial range. The famous Kolmogorov picture (Kolmogorov 1941) corresponds to hydrodynamic turbulence, but, as we discuss further, a qualitatively similar turbulence also develops in magnetized fluids/plasmas. Therefore both turbulence and magnetic fields should be dealt with while addressing the problem of star formation. Direct observations in Milky Way and nearby galaxies provide ample evidence that the star formation happens in magnetized turbulent clouds. Due to the process of turbulent dynamo magnetic energy increases
fast and one may argue that the formation of stars at high redshifts, including the first starts in early universe (see Norman, Wilson \& Barton 1980, Abel et al. 2002, Nakamura \& Umemura 2001, Scheicher et al. 2010) could also take place in turbulent magnetized environments. 

\subsection{Strong and weak Alfvenic turbulence}

For the purposes of reconnection diffusion that we describe below, Alfvenic perturbations are vital. Numerical studies in Cho \& Lazarian (2002, 2003) showed that the Alfvenic turbulence develops an independent cascade which is marginally affected
by the fluid compressibility. This observation corresponds to theoretical expectations of the Goldreich \& Sridhar (1995, henceforth GS95) theory that we briefly describe below (see also Lithwick \& Goldreich 2001). In this respect we note that the MHD approximation is widely used to describe the actual magnetized plasma turbulence over scales that are much larger than both the mean free path of the particles and their Larmor radius (see Kulsrud 1983, 2005 and references therein). More generally, the most important incompressible Alfenic part of the plasma motions can described by MHD even below the mean free path but on the scales larger than the Larmor radius (see also \S 4.2). 

While having a long history of ideas, the theory of MHD turbulence has become
testable recently due to the advent of numerical simulations (see Biskamp 2003)
which confirmed (see Cho \& Lazarian 2005 and references therein) the prediction of magnetized Alfv\'enic eddies
being elongated in the direction of magnetic field (see Shebalin, Matthaeus \&
Montgomery 1983, Higdon 1984) and provided results consistent with the
quantitative relations for the degree of eddy elongation obtained  in GS95.

\begin{table*}[t]
\vskip4mm
\centering
\begin{tabular}{lllll}
\multicolumn{5}{c}{{\bf Table 1}} \\
\multicolumn{5}{c}{Regimes and ranges of MHD turbulence} \\
\hline
\hline
Type          & Injection &  Range   & Motion & Ways\\
of MHD turbulence & velocity & of scales & type & of study\\
\hline
Weak & $V_L<V_A$ & $[L, l_{trans}]$ & wave-like & analytical\\
\hline
Strong &  & & & \\
subAlfvenic& $V_L<V_A$ & $[l_{trans}, l_{min}]$ & eddy-like & numerical \\
\hline
Strong &  &  &  & \\
superAlfvenic & $V_L> V_A$ & $[l_A, l_{min}]$ & eddy-like & numerical \\
\hline
& & & \\
\multicolumn{5}{l}{\footnotesize{$L$ and $l_{min}$ are injection and dissipation scales}}\\
\multicolumn{5}{l}{\footnotesize{$l_{trans}$  and $l_{A}$ are given by Eq. (\ref{trans}) and
Eq. (\ref{alf}), respectively.}}\\
\end{tabular}
\end{table*}

The hydrodynamic counterpart of the MHD turbulence theory is the famous Kolmogorov (1941) theory of turbulence. In the latter theory energy is injected at large scales, creating large eddies which correspond to large $Re$ numbers and therefore do not dissipate energy through viscosity\footnote{Reynolds number $Re\equiv L_fV/\nu=(V/L_f)/(\nu/L^2_f)$ which is the ratio of an eddy turnover rate $\tau^{-1}_{eddy}=V/L_f$ and the viscous dissipation rate $\tau_{dis}^{-1}=\eta/L^2_f$. Therefore large $Re$ correspond to negligible viscous dissipation of large eddies over the cascading time $\tau_{casc}$ which is equal to $\tau_{eddy}$ in Kolmogorov turbulence.} but transfer energy to smaller eddies. The process continues untill the cascade reaches the eddies that are small enough to dissipate energy over eddy turnover time. In the absence of compressibility the hydrodynamic cascade of energy is $\sim v^2_l/\tau_{casc, l} =const$, where $v_l$ is the velocity at the scale $l$ and the cascading time for the eddies of size $l$ is $\tau_{cask, l}\approx l/v_l$. From this the well known relation $v_l\sim l^{1/3}$ follows.

A frequent mental picture that astrophysicists have of the Alfvenic turbulence is based of Alfven waves with wavevectors along the magnetic field. This is not true for the strong Alfvenic turbulence which, similar to its hydrodynamic counterpart, can be described in terms of eddies\footnote{The description in terms of interacting wavepackets or modes is also possible with the corresponding
wavevectors tending to get more and more perpendicular to the magnetic field as the cascade develops.}. However, contrary  to Kolmogorov turbulence, in the presence of dynamically important magnetic field eddies become anisotropic. At the same time, one can imagine eddies mixing magnetic field lines perpendicular to the direction of magnetic field. For the latter eddies the original Kolmogorov treatment is applicable resulting in perpendicular motions scaling as $v_l\sim \l_{\bot}^{1/3}$, where $l_{\bot}$ denotes eddy scales measured perpendicular to magnetic field. These mixing motions induce Alfvenic perturbations that determine the parallel size of the magnetized eddy.  The key stone of the GS95 theory is {\it critical balance}, i.e. the equality of the eddy turnover time $l_{\bot}/v_l$ and the period of the corresponding Alfven wave $\sim l_{\|}/V_A$, where $l_{\|}$ is the parallel eddy scale and $V_A$ is the Alfven velocity. Making use of the earlier expression for $v_l$ one can easily obtain $l_{\|}\sim l_{\bot}^{2/3}$, which reflects the tendency of eddies to become more and more elongated as the energy cascades to smaller scales (see Beresnyak, Lazarian \& Cho 2005).

It is important to stress that the scales $l_{\bot}$ and $l_{\|}$ are measured in respect to the system of reference related to the direction of the local magnetic field "seen" by the eddy. This notion was not present in the original formulation of the GS95 theory and was added to it in LV99. The local system of reference was later used in numerical studies in Cho \& Vishniac (2000), Maron \& Goldreich (2001), Cho, Lazarian \& Vishniac (2002) testing GS95 theory. In terms of mixing motions, it is rather obvious that the free Kolmogorov-type mixing is possible only in respect to the local magnetic field of the eddy rather than the mean magnetic field of the flow. 

While the arguments above are far from being rigorous (see more in a review by Cho \& Lazarian 2005), they correctly reproduce the basic scalings of magnetized turbulence when the velocity equal to $V_A$ at the injection scale $L$. The most serious argument against the picture is the ability of eddies to perform mixing motions perpendicular to magnetic field. Jumping ahead of our story, we can mention that this ability is related to the ability of magnetic field lines to reconnect fast, i.e. at the rate independent of the fluid resistivity (see more in \S 5.1).  

GS95 theory assumes the isotropic injection of energy at scale $L$ and the injection velocity equal to the Alfv\'en velocity in
the fluid $V_A$, i.e. the Alfv\'en Mach number $M_A\equiv (V_L/V_A)=1$, where $V_L$ is the injection velocity. Thus it provides the description
of transAlfvenic turbulence. This model was later generalized
for both subAlfvenic, i.e. $M_A<1$, and superAlfvenic, i.e. $M_A>1$, cases (see Lazarian \& Vishniac 1999 and Lazarian 2006, respectively; see also
Table~1).  Indeed, if $M_A>1$, instead of the driving scale $L$ for  one can use another scale,
namely $l_A$ (see Eq. (\ref{alf}), which is the scale at
which the turbulent velocity equals to $V_A$.  For $M_A\gg 1$
magnetic fields are not dynamically important at the largest scales and the
turbulence at those scales follows the isotropic
 Kolmogorov cascade $v_l\sim l^{1/3}$ over the
range of scales $[L, l_A]$. 
At the same time, if $M_A<1$, the turbulence obeys GS95 scaling (also called ``strong''
MHD turbulence) not from the scale $L$, but from a smaller scale $l_{trans}$ given by Eq. (\ref{trans}), while in the range $[L, l_{trans}]$ the turbulence is ``weak''. We discuss
more superAlfvenic and subAlfvenic turbulence in \S 5.1. 

The properties of weak and strong turbulence are rather different. Weak turbulence is wave-like turbulence with wave packets undergoing many collisions before transferring energy to small scales. It corresponds well to the mental picture of turbulence of weakly interacting Alfvenic waves that used to dominate astrophysics textbooks. Weak turbulence, unlike the strong one, allows an exact analytical treatment (Gaultier et al. 2000). On the contrary, the strong turbulence is eddy-like with cascading happening similar to Kolmogorov turbulence within roughly an eddy turnover time. The strong interactions between wave packets 
prevent the use of perturbative approach and do not allow exact derivations. It were the numerical experiments that proved the predicted scalings for incompressible MHD turbulence (see Cho \& Vishniac 2000, Maron \& Goldreich 2001, Cho et al. 2002, Beresnyak \& Lazarian 2010, Beresnyak 2011) and for the Alfvenic component of the compressible MHD turbulence\footnote{For compressible MHD turbulence simulations in Beresnyak et al. (2005) and Kowal,
Lazarian \& Beresnyak (2007) demonstrated that the density spectrum becomes more shallow and isotropic as the Mach number increases.} (Cho \& Lazarian 2002, 2003, Kowal \& Lazarian 2010). 

One also should keep in mind that the notion "strong" should not be associated with the amplitude of turbulent motions but only with the strength of the non-linear interaction. As the weak turbulence evolves, the interactions of wave packets {\it get stronger} making the turbulence strong. In this case, the amplitude of the perturbations can be very small.

While there are ongoing debates whether the original GS95 theory must be modified to better describe MHD turbulence, we believe that, first of all, we do not have compelling evidence that GS95 is not adequate\footnote{Recent work by Beresnyak \& Lazarian (2009a, 2010) shows that present day numerical simulations are unable to reveal the actual inertial range of MHD turbulence making the discussions of the discrepancies of the numerically measured spectrum and the GS95 predictions rather premature. In addition, new higher resolution simulations by Beresnyak (2011) reveal the predicted $-5/3$ spectral slope.}. Moreover, the proposed additions to the GS95 model do not change the nature of the physical processes that we discuss below.

\section{Magnetic reconnection of turbulent magnetic field}

In what follows we describe the model of reconnection of magnetic field proposed in LV99.   

\subsection{Turbulence in reconnection zone}

Magnetic reconnection is a fundamental process that violates the frozen-in state of magnetic flux. 
We would like to stress that we are discussing the case of dynamically important magnetic field, including the case of a weakly turbulent magnetic field. The case of a weak magnetic field which can be easily stretched and bent by turbulence at any scale up to the dissipation one is rather trivial and of little astrophysical significance for star formation (see more in \S 5.1). Indeed, at sufficiently small scales magnetic fields get dynamically important even for superAlfvenic turbulence (see Eq. (\ref{alf})).  

\begin{figure}
\centering
  \includegraphics[height=.30\textheight]{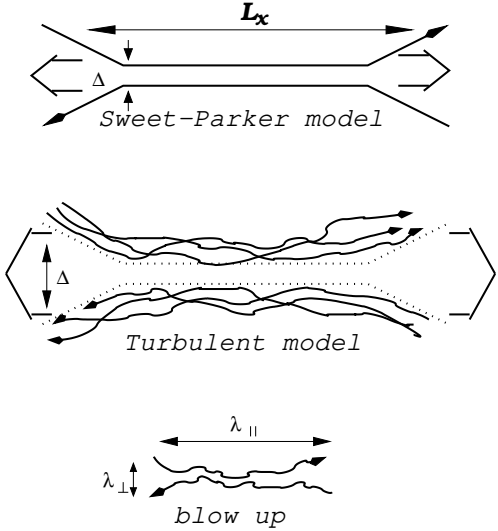}
  \caption{{\it Upper panel}: Sweet-Parker reconnection. $\Delta$ is limited by resistivity and small. {\it Middle panel}: reconnection according to LV99 model. $\Delta$ is determined by turbulent field wandering and can be large. {\it Lower panel}: magnetic field reconnect over small scales. From Lazarian, Vishniac \& Cho (2004).}
 \label{LV}
\end{figure}

Within the picture of eddies mixing perpendicular to the local magnetic field (see \S 3.2), it is suggestive that magnetized eddies can induce turbulent diffusion similar to ordinary hydrodynamic eddies. This is a rather counter-intuitive notion in view of the well-entrenched idea of flux being frozen in to astrophysical fluids. As it is explained in ELV11 the frozen-in condition is not a good approximation for turbulent fluids\footnote{Formal mathematical arguments on how and why the frozen-in condition fails may be found in Eyink (2011) (see also Eyink 2007, 2009).}. The violation of the frozen in condition  also follows from LV99 model  of reconnection (see discussion in Vishniac \& Lazarian 1999).

A picture of two flux tubes of different directions which get into contact in 3D space is a generic framework to
describe magnetic reconnection. The upper panel of Figure \ref{LV} illustrates why reconnection is so slow in the textbook Sweet-Parker model. Indeed, the model considers magnetic fields that are laminar and therefore the frozen-in condition for magnetic field is violated only over a thin layer dominated by plasma resistivity. The scales over which the resistive diffusion is important are microscopic and therefore the layer is very thin, i.e. $\Delta\ll L_x$, where $L_x$ is the scale at which magnetic flux tubes come into contact. The latter is of the order of the diameter of the flux tubes and typically very large for astrophysical conditions. During the process of magnetic reconnection all the plasma and the shared magnetic flux\footnote{Figure \ref{LV}  presents a cross section of the 3D reconnection layer. A shared component of magnetic field is present in the generic 3D configurations of reconnecting magnetic flux tubes.} arriving over an astrophysical scale $L_x$ should be ejected through a microscopic slot of thickness $\Delta$. As the ejection velocity of magnetized plasmas is limited by Alfven velocity $V_A$, this automatically means that the velocity in the vertical direction, which is reconnection velocity, is much smaller than $V_A$.

Being more quantitative, one can write
\begin{equation}  
v_{rec}= V_A \frac{\Delta}{L_x}, 
\label{vrec} 
\end{equation}
meaning that $v_{rec}\ll V_A$ if $\Delta \ll L_x$.
There are different ways to derive Sweet-Parker formulae for the reconnection rate (see Parker 1979). One way is to 
consider the Ohmic diffusion of magnetic field lines (see ELV11). The mean-square 
vertical distance that a magnetic field-line can diffuse by resistivity in time $t$ is 
\begin{equation}
\langle y^2(t)\rangle \sim \lambda t, 
\label{diff-dist} 
\end{equation}
where $\lambda=\eta c^2/4\pi$ is Ohmic diffusivity.
The field lines are advected out of the sides of the 
reconnection layer of length $L_x$ at a velocity of order $V_A.$ Thus, the time that the lines can 
spend in the resistive layer is the Alfv\'en crossing time $t_A=L_x/V_A.$ Thus, field lines can only 
be merged that are separated by a distance 
\begin{equation}
\Delta = \sqrt{\langle y^2(t_A)\rangle} \sim \sqrt{\lambda t_A} = L_x/\sqrt{S},
\label{Delta} 
\end{equation}
where $S$ is Lundquist number,
\begin{equation}
S=L_x V_A/\lambda.
\label{Lun}
\end{equation}
Combining Eqs. (\ref{vrec}) and (\ref{Delta}) one gets the famous Sweet-Parker reconnection rate, 
\begin{equation}
v_{rec, SP}=V_A/\sqrt{S}.
\label{SP}
\end{equation}

The LV99 model generalizes the Sweet-Parker one by accounting for the existence of magnetic field line stochasticity (Figure \ref{LV} [lower panels]).  The shown turbulence is subAlfvenic and the mean field is clearly defined. At the same time turbulence induces magnetic field wandering. This wandering was quantified in LV99 and it depends
on the intensity of turbulence. The vertical extend of wandering of magnetic field lines that at any point get into contact with the field of the other flux tube was identified in LV99 with the width of the outflow region. In other words, the LV99 model of reconnection makes use of the fact that in the presence of magnetic field wandering, which is a characteristic feature of magnetized turbulence in 3D, the outflow is no more constrained by the narrow resistive layer, but happens through a much wider area $\Delta$ defined by wandering magnetic field lines. An
important consequence of this is that as turbulence
amplitude increases, the outflow region and therefore reconnection rate also increases, which entails the 
ability of reconnection to change its rate depending on the level of turbulence. The latter is important both
for understanding the dynamics of magnetic field in turbulent flow and for explaining flaring reconnection events, e.g. solar flares.  

We should note that the magnetic field wandering is mostly due to Alfvenic turbulence\footnote{As discussed in LV99 and in more details in ELV11 the magnetic field wandering, turbulence and magnetic reconnection are very tightly related concepts. Without magnetic reconnection, properties of magnetic turbulence and magnetic field wandering would be very different. For instance, in the absence of fast reconnection, the formation of magnetic knots arising if magnetic fields were not
able to reconnect would destroy the self-similar cascade of Alfvenic turbulence. The rates predicted by LV99 are
the rates required to make Goldreich-Sridhar model of turbulence self-consistent.} and the corresponding
expressions for $\Delta$ arising from the field wandering were obtained in LV99. An alternative derivation of the 
$\Delta$ was obtained in analogy with the Sweet-Parker derivation above in ELV11 appealing to the concept
of Richardson (1926) diffusion.

Richardson diffusion (see Kupiainen et al. 2003) implies the mean squared separation of particles
\begin{equation}
\langle |x_1(t)-x_2(t)|^2 \rangle\approx \epsilon t^3,
\label{Rich}
\end{equation}
 where $t$ is time, $\epsilon$ is the energy cascading rate and $\langleÉ\rangle$ denote an ensemble averaging. For subAlfvenic turbulence $\epsilon\approx u_L^4/(V_A L)$,
where $u_L$ is the injection velocity and $L$ is an injection scale (see LV99) and therefore analogously to Eq. (\ref{Delta}) one can write
\begin{equation}
\Delta\approx \sqrt{\epsilon t_A^3}\approx L_x(L_x/L)^{1/2}M_A^2
\label{D2}
\end{equation}
where it is assumed that $L_x<L$. Combining Eqs. (\ref{vrec}) and (\ref{D2})
one recovers the LV99 expression for the rate of magnetic reconnection (see also ELV11)
\begin{equation}
v_{rec, LV99}\approx V_A (L_x/L)^{1/2}M_A^2.
\label{LV99}
\end{equation}
in the limit of $L_x<L$. Analogous considerations allow to recover the LV99 expression for $L_x>L$, which differs 
from Eq.~(\ref{LV99}) by the change of the power $1/2$ to $-1/2$.  

It is important to stress that Richardson diffusion ultimately leads to the diffusion over the entire width of large scale eddies once the plasma has moved the length of one such eddy. The precise scaling exponents for the turbulent cascade does not affect this result, and all of the alternative scalings considered in LV99 yield the same behavior.

\begin{figure}
\center
\includegraphics[width=0.9\columnwidth]{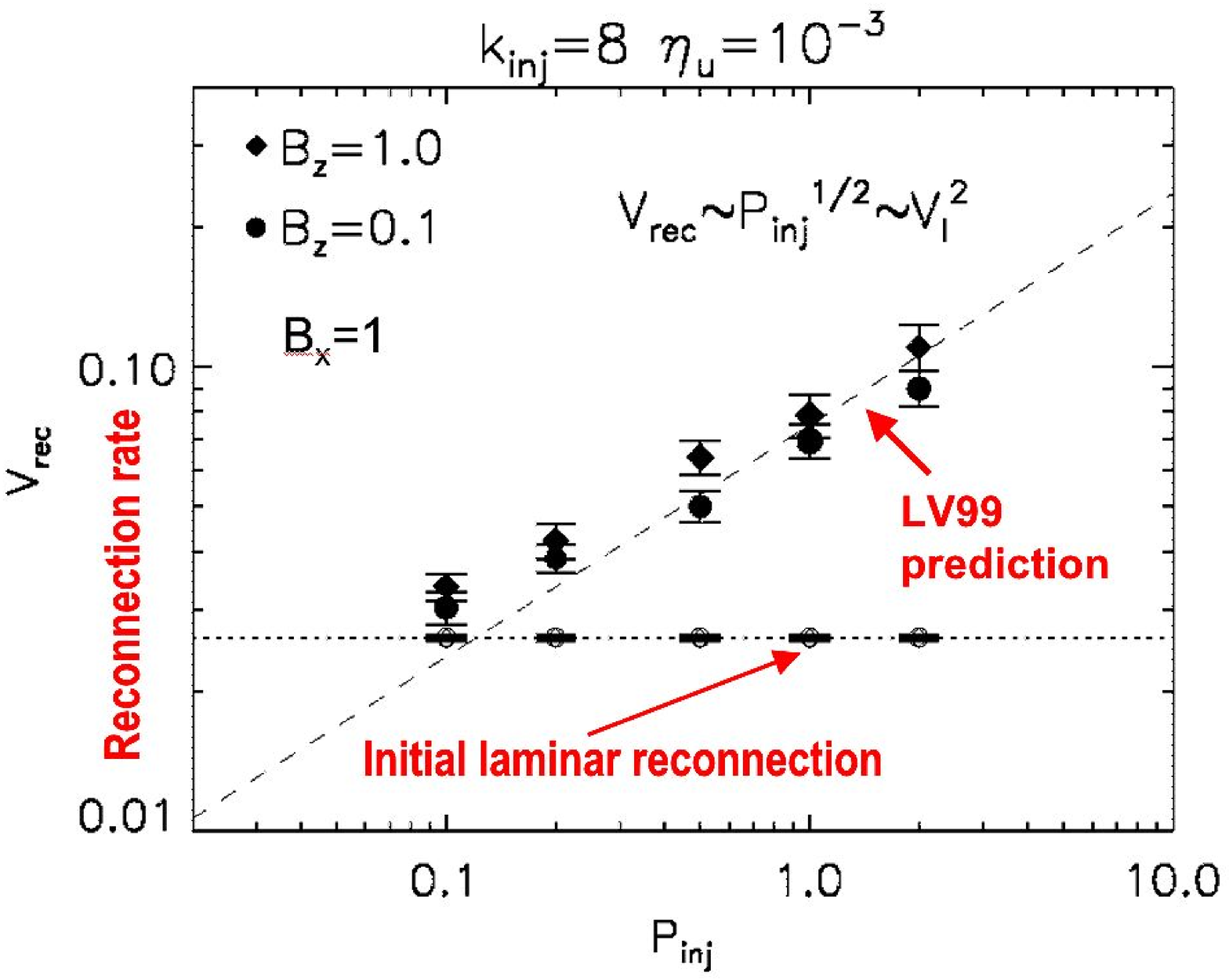}
\caption{Reconnection speed versus input power for the driven turbulence.  We show the reconnection speed plotted against the input power for an injection wavenumber equal to 8 (i.e. a wavelength equal to one eighth of the box size) and a resistivity $\nu_u$.  The dashed line is a fit to the predicted dependence of  $P^{1/2}$ (see eq. (3)).  The horizontal line shows the laminar reconnection rates for each of the simulations before the turbulent forcing started.  Here the   uncertainty in the time averages are indicated by the size of the symbols and the variances are shown by the error bars. Modified from Kowal et al. (2009). 
\label{pow_dep}}
\end{figure}

The predictions of the turbulent reconnection rates in LV99 were successfully tested 3D numerical simualtions in Kowal et al. (2009) (see also Lazarian et al. 2010 for an example of higher resolution runs). In Figure~\ref{pow_dep} we see the results for varying amounts of input power, for fixed resistivity and injection scale as well as for the case of no turbulence at all.  The line drawn through the simulation points is for the predicted scaling with the square root of the input power. The agreement between equation (\ref{LV99}) and Figure~\ref{pow_dep} is encouraging. Similarly the dependences of the reconnection rate on the injection scale and on the
Ohmic and anomalous resistivity were successfully tested. 

The testing of LV99 predictions provides more confidence to the theory and stimulates to think of its applications,
e.g. for star formation. One should keep in mind that the LV99 model assumes that the magnetic field flux tubes can come at arbitrary angle, which corresponds to the existence of shared or guide field within the reconnection layer\footnote{The model in LV99 is three dimensional, and it is not clear to what extend it can be applied to 2D turbulence (see discussion in ELV11 and references therein). However, the cases of pure 2D reconnection and 2D turbulence are of little practical importance.}.

\subsection{LV99 Reconnection and Plasma effects}

For years plasma effects have been considered essential for fast magnetic reconnection (see Shay et al. 1998, Daughton et al.  2006, 2008).  On
the contrary, LV99 makes use of the MHD approximation. To issue of the justification of  the MHD description of plasmas in the LV99 model was recently revisited in ELV11.  One can think of three relevant
length-scales: the ion gyroradius $\rho_i,$ the ion mean-free-path length 
$\ell_{mfp,i}$, and the scale $L_{s}$ of large-scale variation of magnetic 
and velocity fields. Astrophysical plasmas can be ``strongly collisional''
if $\ell_{mfp,i}\ll \rho_i,$ and can be described as fluids. 
The interiors of stars and accretion disks present examples of such plasmas.  Another case is 
 ``weakly collisional''  $\ell_{mfp,i}\gg \rho_i$ plasmas. The ratio of the mean free path to the gyroscale
 \begin{equation}
 \frac{\ell_{mfp,i}}{\rho_i}\propto \frac{\Lambda}{\ln\Lambda}\frac{v_A}{c}, 
\label{lmfp-rho} 
\end{equation}
follows from the expression for the Coulomb collision frequency (see 
Fitzpatrick 2011, eq.(1.25)), where $\Lambda=4\pi n\lambda_D^3$ is the plasma parameter,
or number of particles within the Debye screening sphere. Hot and rarified astrophysical plasmas are  
weakly coupled which entails $\Lambda$ being large, for instance, of the order of $10^9$ or
more for the warm component of the interstellar medium or solar wind (see Table~1 in ELV11). 
For such ratio the expansion over small ion gyroradius results in ``kinetic MHD equations''.
Those differ from the standard MHD by having anisotropic pressure tensor (see more 
discussion in Kowal et al. 2011a and references therein).

In addition, plasmas that are not strongly collisional can be divided into two subclasses: ``collisionless'' 
plasmas for which the mean free pass is larger than  the largest scales of interest
$\ell_{mfp,i}\gg L_s,$, and ``weakly collisional'' plasmas 
for which the opposite is true, i.e.
$L_s\gg \ell_{mfp,i}.$ In the latter case the``kinetic MHD'' description allows further reduction in complexity 
at  scales larger the mean free path $\ell_{mfp,i}$. This, as discussed in ELV11, reproduces a fully hydrodynamic MHD description {\it at 
those scales}.  For instance, the warm ionized ISM is ``weakly collisional'', 
while the solar wind interaction with the magnetosphere is ``collisionless.''

Additional simplifications are possible when (a) turbulent fluctuations 
are small compared to the mean magnetic field,  (b) have length-scales parallel to the mean field much larger than 
perpendicular length-scales, and © have frequencies low compared to the ion cyclotron frequency. This set of assumptions (a), (b) and (c)
is adopted in the GS95 theory and for
``gyrokinetic approximation''  (see Schekochihin et al 2007). At length-scales, i.e. at scales much larger than $\rho_i$, an important simplification takes place. At those scales incompressible
shear-Alfven wave modes have get independent from the compressive modes and described 
by the simple ``reduced MHD'' (RMHD) equations (see GS95, Cho \& Lazarian 2003). This fact is of major importance for the LV99 justifying the use of the analysis based on an incompressible MHD fluid model. 

\begin{figure}
\centering
  \includegraphics[height=.20\textheight]{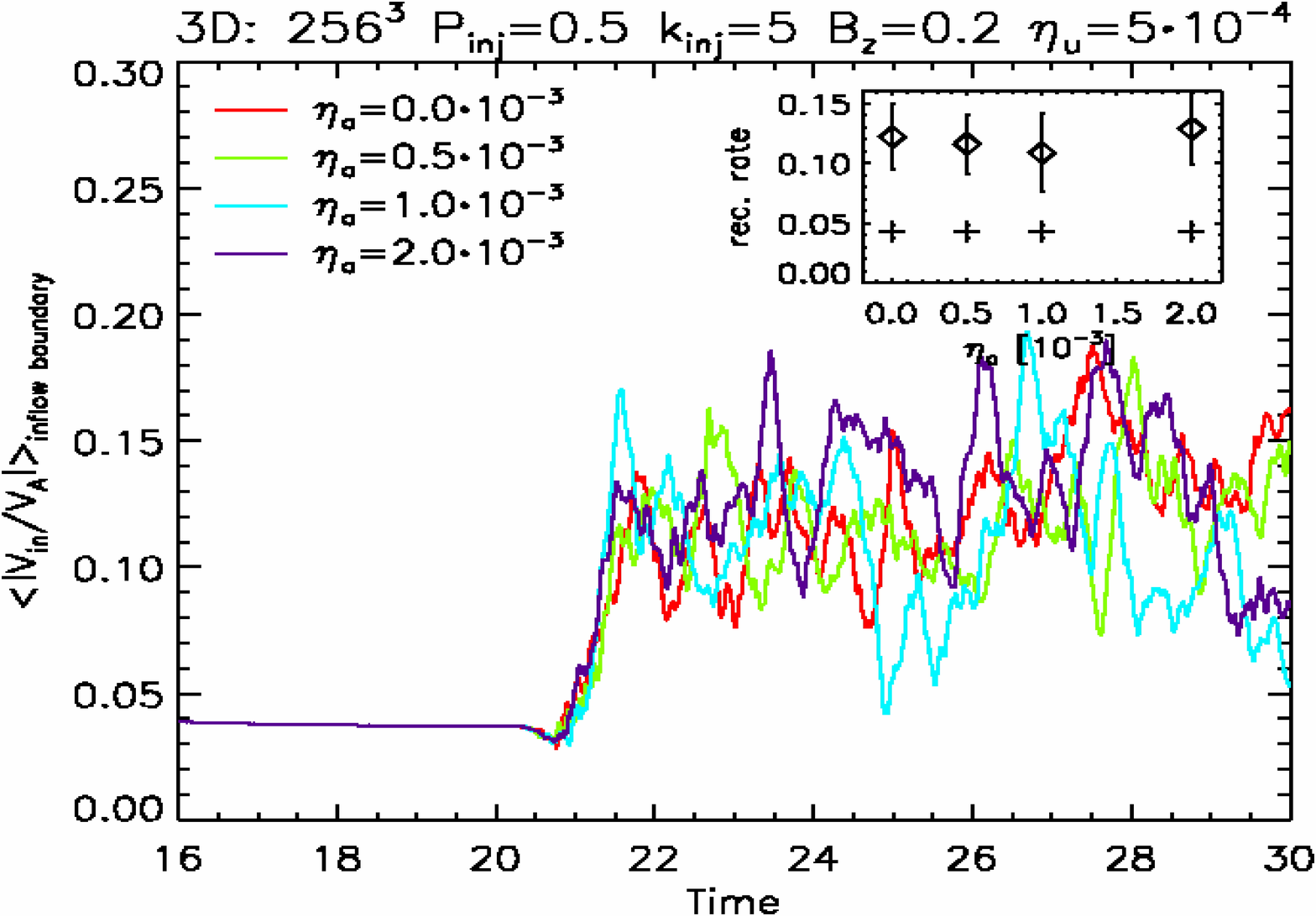}
  \includegraphics[height=.20\textheight]{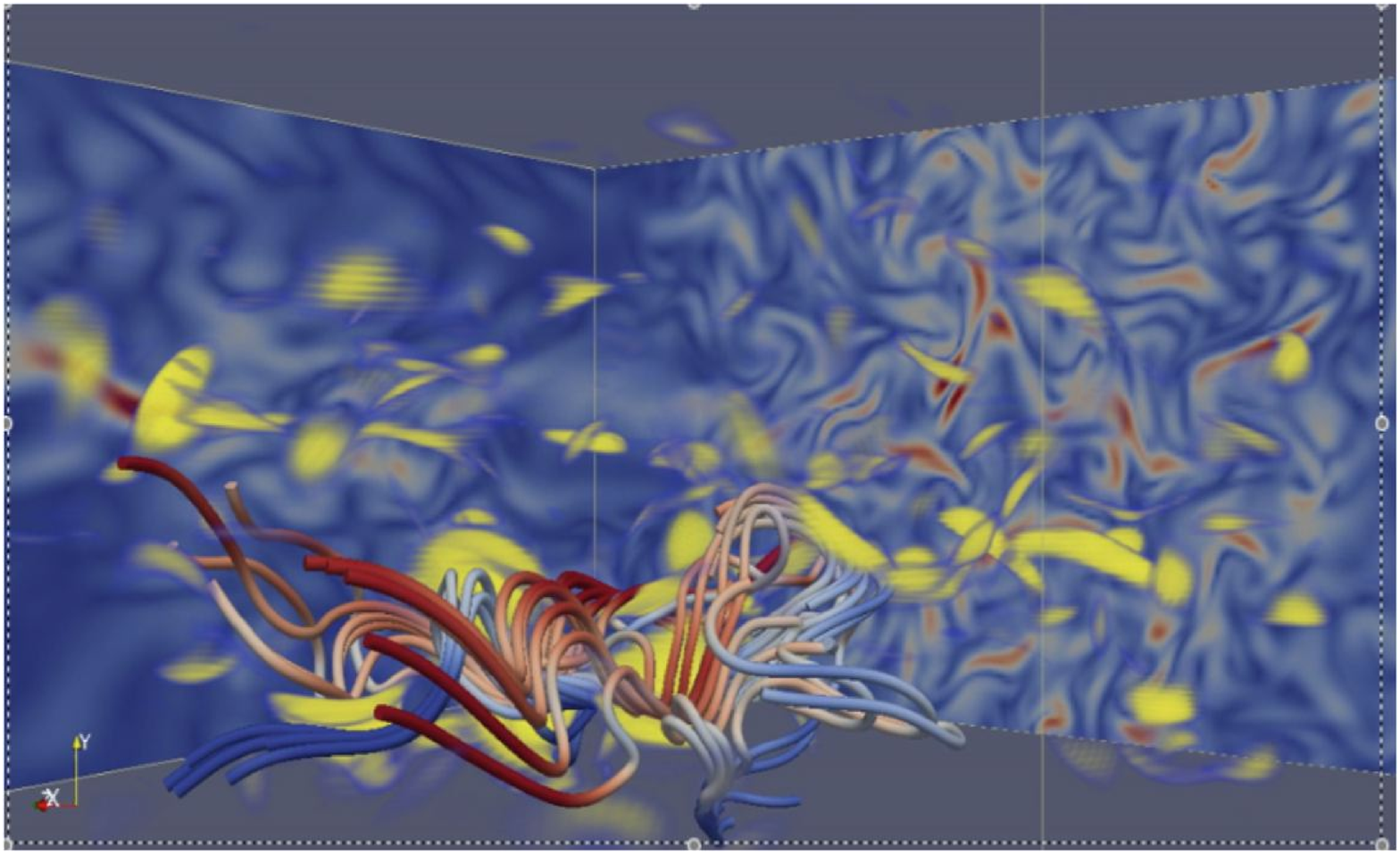}
  \caption{ {\it Upper plot}:Reconnection rate in the presence of anomalous resistivity and turbulence. Changes of the anomalous resistivity do not change the reconnection rate. {\it Lower plot}: Visualization of reconnection region. Turbulence is weakly driven and
the big changes in magnetic field lines are due to reconnection. Plotted with data in Kowal et al. (2009).}
 \label{kov2}
\end{figure}

For most astrophysical problems, including the problem of star formation, one has deal with the reconnection at length scales much
larger than ion gyroradius, i.e. $L_s\gg \rho_i$. For the solar wind at in the Earth magnetosphere where some of the reconnection
measurement have been done (Retino et al. 2007), the ion gyroradius $\rho_i\approx 6\times 10^6$~cm, i.e. comparable with $L_s$. 
In these circumstances, for the span of scales from $\rho_i$ to the electron gyroradius $\rho_e$ the plasmas is described by an
ion kinetic equation and electron "reduced MHD" (ERMHD) equations for kinetic Alfven modes (see Schekochihin et al. 2009). This is
the domain of Hall physics, with ion and electrons moving very differently and "Hall reconnection" being expected 
(see Uzdensky \& Kulsrud 2006). Magnetic field wandering still can take place due to the electron fluid (see Daughton et al. 2011)
but the nature of the turbulent cascade is different (see Cho \& Lazarian 2004, 2009, Schekochihin et al. 2007, 2009). We hope
that future research will clarify to what extend features of the LV99 model carries over to this very different regime of reconnection
in magnetosphere.   

Within the LV99 model, the reconnection rate is determined by large scale magnetic wandering (see Figure \ref{LV} for a pictorial representation
and Figure \ref{wand} for the analysis of simulations), while small scale plasma effects  are irrelevant for
the global reconnection\footnote{It is shown in LV99 that if the reconnection is calculated assuming that the
bottle neck is due to Ohmic resistivity, the reconnection rate gets much larger than the Alfven speed. This
is due to the effect of 3D turbulent magnetic fluxes getting in contact over many independent patches. The
total cumulative rate of reconnection "cutting magnetic field lines" becomes very large and it is the outflow of
magnetized plasma from the reconnection region that enters to limit the overall reconnection speed. Naturally,
increasing the local reconnection speed of magnetic patches either due to plasma effects or, alternatively, numerical effects in computer simulations does not increase the reconnection speed of the large scale turbulent
magnetic fluxes.}  at least in the fully ionized plasma (see Lazarian et al. 2004). This conclusion
is supported by simulations in Kowal et al. (2009) where plasma effect were simulated by using anomalous
resistivity, i.e. the resistivity that depends on the value of the current. Figure \ref{kov2} shows that substantial variations of the anomalous resistivity do not change the resulting reconnection rates. Note that the numerical
effects produce also a sort of anomalous resistivity on the scales comparable to the grid size. Therefore our results testify that, in the presence of turbulence, numerical effects do not dominate simulations in terms
of reconnection. We should mention that although plasma effects do not change the global reconnection
rate, they can be important for other processes, e.g. the acceleration of electrons (see \S 10.1).

\subsection{Alternative views on fast reconnection}

Alternative models of magnetic reconnection appeal to different physics to overcome the limitations of the Sweet-Parker model. In the Petschek (1964) model of reconnection the reconnection layer opens up to enable the
outflow which thickness does not depend on resistivity. There the extend of the current sheet gets of the order of microscopic $\Delta$
and therefore Eq. (\ref{vrec}) provides $v_{rec}\approx V_A$. For years this had been considered the only way of make magnetic
reconnection fast. To realize this idea inhomogeneous resistivity, e.g. anomalous resisitivity associated with plasma effects, is required (see Shay \& Drake 1998, Shay et al. 1998, 1999, Bhattacharjee et al. 2005, Cassak et al. 2006, 2008, 2009). However, for turbulent plasmas, the effects arising from modifying the local reconnection events by introducing anomalous
resistivity are negligible as confirmed e.g. in Kowal et al. (2009).  Other effects, e.g. formation and ejection of plasmoids (see Shibata \& Tanuma 2001, Lourreiro et al. 2007, Uzdensky et al. 2010, Huang et al. 2011) may be important for initially laminar environments 
and contribute to the onset of reconnection. However, for the problems of star formation, it is more important to account for
the pre-existing turbulence for which LV99 model is directly applicable.

Turbulence is known to make accelerated diffusive processes. Therefore it is not surprising that it had been appealed as the way of speeding up the reconnection prior to LV99 study. Nevertheless, LV99 model is radically different from its predecessors which also appealed 
to the effects of turbulence. For instance, unlike Speiser (1970) and 
Jacobson (1984) the model does not appeal to changes of the microscopic properties of the plasma. The closest in its
approach to LV99 
among papers dealing with effects of turbulence was the work of  Matthaeus \& Lamkin (1985, 1986) who studied the problem numerically in 2D MHD and 
who suggested that magnetic reconnection 
may be fast due to a number of  turbulence effects, e.g. multiple X points and turbulent EMF.
 However, the physics of reconnection discussed in Matthaeus \& Lamkin (1985, 1986) is very different, as they did not realize the key role played by magnetic field-line 
wandering, which is the corner stone idea of the LV99 model. They did not obtain a quantitative prediction for the reconnection rate, as did LV99.

A number of earlier papers may be seen as {\it indirect} evidence of fast reconnection in turbulent
fluids. For instance, a study of tearing instability of current sheets in the presence 
of background 2D turbulence and the formation of large-scale, long-lived
magnetic islands has been performed in \cite{Politanoetal89}. They present 
evidence for ``fast energy dissipation'' in 2D MHD turbulence and show that
this result does not change as they change the resolution. A more recent study of 
of \cite{MininniPouquet09} also provides evidence for ``fast dissipation'' but in 
3D MHD turbulence.  This phenomenon is consistent with the idea
of fast reconnection, but cannot be treated as a direct evidence of the process. 
It is very clear that {\it fast energy dissipation} and {\it fast magnetic reconnection} are distinclty different  
physical processes. In addition, a paper by Galsgaard and Nordlund, 
\cite{GalsgaardNordlund97b}, might also be interpreted as 
providing indirect evidence for fast reconnection.  The authors noted that in their 
simulations they could not produce highly twisted magnetic fields. One of the 
interpretations of this finding could be the relaxation of magnetic field via reconnection. 
In this case, this observations could be related to the numerical finding
of \cite{LapentaBettarini11} which shows that reconnecting magnetic configurations 
spontaneously get chaotic and dissipate, which in its turn may be related 
to the predictions in LV99 (see more in Lapenta \& Lazarian 2011). However, in view of the uncertainties of the
numerical studies, it is difficult to be confident of this connection.

LV99 model deals with balanced turbulence where the energy flows in opposite directions are equal. If the latter is not true,
MHD turbulence is imbalanced, or has non-zero cross helicity. Solar wind presents an example of system with imbalanced turbulence.
There have been recent attempts to study reconnection in systems with a flow and imbalanced turbulence (Yokoi \& Hoshino 2011). We
feel that to obtain quantitative predictions one needs to use the scaling properties of the turbulence as it is done in LV99. At the
same time the theory of strong imbalanced MHD turbulence is still controversial.  Among the existing theories of imbalanced 
turbulence (see \cite{LithwickGoldreich07, BeresnyakLazarian08, Chandran08, PerezBoldyrev09}, all, but 
\cite{BeresnyakLazarian08} seem to contradict to numerical testing in Beresnyak \& Lazarian (2009b). We defer the discussion of
reconnection in imbalanced turbulence to future publications\footnote{We expect the effect of field wandering to play the crucial role
for the reconnection with non-zero cross-helicity. This wandering should be determined using the theory of strong imbalanced MHD
turbulence, for instance, by one in Beresnyak \& Lazarian (2008) if higher resolution and longer averaging future testings confirm it. We also expect that similarly as a successful model of imbalanced MHD turbulence must produce GS95 scaling for the case of zero cross helicity,
the reconnection in flows with imbalanced turbulence should converge to LV99 predictions for the zero cross helicity. For very high cross helicity field wandering may be small and other, e.g. plasma effects, become important.}.

It is important to stress that while possible reconnection schemes may be numerous, the {\it generic astrophysical
 reconnection model} should satisfy several constraints. A fundamental consideration for such a model is that they must explain fast reconnection in collisional
and collisionless plasmas.  At the same time, to explain flares, it should be possible for reconnection to be delayed for significant
amounts of time. The reconnection model should be able to operate in the turbulent environment
as astrophysical media are turbulent. As far as we know, only LV99 model satisfies to all of these requirements.

For instance,  LV99 model explains the accumulation of flux in highly magnetized plasmas if the level of turbulence is low. It is possible that tearing mechanisms may provide original perturbations stimulating the development of turbulence and reconnection that it induces. As the outflow within the reconnection region gets turbulent, it induces more of
field wandering of the reconnecting fluxes and therefore higher reconnection speed (see more in Lazarian \& Vishniac 2009). This introduces the positive feedback which results in
a flare. LV99 predicted that turbulence from neighboring regions can also ignite reconnection and the observation of the initiation of flares by incoming Alfvenic waves was reported by Sych et al. (2009). 

We would like to stress, that magnetic turbulence and reconnection are intrinsically connected. Therefore there is no  "magic"
reconnection rate, e.g. $0.1 V_A$, that satisfies all the requirements\footnote{The latter was a sort of Holy Grail for many researchers
studying reconnection.} For instance, to avoid formation
of magnetic knots turbulent eddies should be able to reform their magnetic field structure over
the time scales of their turnover. For transAlfvenic turbulence this means the reconnection rate $\sim V_A$. This stringent constraint is satisfied by the LV99 model (see \S 5.1). 

\section{Reconnection diffusion concept}

\subsection{Diffusion in magnetized turbulent fluid}

The exact treatment of diffusion in turbulent fluid is rather complicated and it requires dealing with 
magnetic reconnection as a part of the process. Therefore we first discuss an illustrative example with 
pure hydro turbulence.  

{\it Hydrodynamic Diffusion}\\
To illustrate analytical approaches to diffusion in a turbulent fluid one can consider first incompressible unmagnetized fluid where the velocity $U$ is decomposed into a regular part $V$ and a fluctuation $v$. Averaging the Navier-Stockes equations one gets
\begin{equation}
\frac{\partial V_i}{\partial t}+V_j\frac{V_i}{\partial x_j}=-\frac{1}{\rho}\frac{\partial P}{\partial x_j}+\frac{\mu}{\rho}\frac{\partial^2V_i}{\partial x_j^2}-\frac{\partial}{\partial x_j}\langle v_i v_j\rangle
\end{equation}
where $P$ is the average pressure, $\langle ..\rangle$ denote averaging procedure and the indices indicate vectors and the standard summation convention. The term $\langle v_i v_j\rangle$ is not specified by the equation and its approximation involves the different "closures" (see Monin \& Yaglom 1975 for a discussion of various hydrodynamic closures). A similar problem emerges in the description of the diffusion induced by turbulence.

Consider as an example the diffusion of a scalar quantity $s$, e.g. a passive impurity, e.g. heavy elements in the ISM. If only molecular diffusion were present, the rate of transport of $s$ in $x$-direction would be given by Fick's law, namely, 
\begin{equation}
q=-D\frac{\partial s}{\partial x} 
\label{fick}
\end{equation}
where $D$ is the molecular diffusion coefficient. This law is being modified by the advection of the quantity $s$ by the transport induced by the velocity field. Then one can get the advective - diffusion equation (see Fisher 1979):
\begin{equation}
\frac{\partial s}{\partial t} + U_x \frac{\partial s}{\partial x} + U_y \frac{\partial s}{\partial y}+ U_z \frac{\partial s}{\partial z}=D\left(\frac{\partial^2 s}{\partial x^2} +\frac{\partial^2 s}{\partial y^2} +\frac{\partial^2 s}{\partial z^2} \right)
\label{advec}
\end{equation}

Decomposing the field $s$ into the mean $s_m$ and fluctuating $s_f$ parts one gets after averaging from Eq.~\ref{advec}
\begin{equation}
\frac{\partial s_m}{\partial t}+V_j\frac{\partial s_m}{\partial x_j}=\frac{\partial}{\partial x_j} \left(D \frac{\partial^2 s_m}{\partial x_j} - \langle v_j s_f
\rangle\right)
\label{s-diff}
\end{equation} 
where the last term corresponds to the transport of the field by turbulent fluctuations compared to the Fick's law based on molecular diffusion (see Eq.~\ref{fick}). The frequently used "closure" appeals to "eddy diffusivity" coefficients defined as follows
\begin{equation}
\langle v_j s_f \rangle=-\epsilon_l \frac{\partial s}{\partial x_j} 
\label{closure}
\end{equation}
where, in general, coefficients $\epsilon$ may depend on the direction. For hydrodynamic turbulence $\epsilon_l \sim v_l l$, which results in
Richardson (1926) diffusion coefficient $\epsilon \sim l^{4/3}$ if velocity field is Kolmogorov, i.e.  $v_l\sim l^{1/3}$. For the diffusion at the largest scales induced by turbulence with the injection scale $L$ and velocity $V_L$
\begin{equation}
\kappa_{hydro}\sim L V_L
\label{hydro}
\end{equation}
corresponds to the maximal diffusivity of unmagnetized turbulent fluid. The above treatment can approximate the dynamics
of fluid perpendicular to the local magnetic field, provided that the reconnection is fast. This is the issue that we address
below.

{\it Diffusion in Magnetized Fluid}\\
The introduction of magnetic field complicates the process of diffusion as it forces one to account for the back reaction of magnetic field. The latter was a hotly debated subject in the dynamo theory (see Parker 1979). For infinitesimally weak magnetic field which backreaction is negligible the turbulent eddies bend magnetic field lines over all scales up to the Ohmic dissipation one and the disparity of scales present in the problem of reconnection described in \S 4.1 does not emerge. This way of reasoning resulted in the concept of {\it magnetic turbulent diffusivity} within kinematic dynamo according to which the diffusion of magnetic field should is similar to the diffusion of passive scalar in hydrodynamic turbulence. However, this regime of dynamically unimportant magnetic field presents a highly unrealistic case of marginal astrophysical importance\footnote{In the case of a dynamically unimportant field the magnetic dissipation and reconnection happens on the scales of the Ohmic diffusion scale and the effects of magnetic field on the turbulent cascade are negligible. However, turbulent motions transfer an appreciable portion of the cascading energy into magnetic energy (see Cho et al. 2009; also \S 3.1). As a result, the state of intensive turbulence with negligible magnetic field is short-lived.}.

The problem of diffusion of magnetized fluid when magnetic fields are dynamically important demands that the issue of
magnetic reconnection is to be addressed. Otherwise, the existence of eddies is questionable. Mixing motions in MHD turbulence require that reconnection events in MHD turbulence should happen through every eddy turnover.  This is, however, what the LV99 model predicts.  Indeed, for small scales magnetic field lines are nearly parallel and, when they intersect, the pressure gradient is not $V_A^2/l_{\|}$ but rather $(l_{\bot}^2/l_{\|}^3) V_A^2$, since only the energy of the component of the magnetic field that is not shared is available to drive the outflow. On the other hand, the characteristic length contraction of a given field line due to reconnection between adjacent eddies is $l_{\bot}^2/l_{\|}$. This gives an effective ejection rate of $V_A/l_{\|}$. Since the width of the diffusion layer over the length $l_{\|}$ is $l_{\bot}$ Eq.(\ref{LV99}) should be replaced by $V_{R}\approx V_A (l_{\bot}/l_{\|})$, which provides the reconnection rate $V_A/l_{\|}$, which is just the nonlinear cascade rate on the scale $l_{\|}$. This ensures self-consistency of critical balance for strong Alfvenic turbulence in highly conducting fluids (LV99). In fact, if not for the LV99 reconnection the buildup of unresolved magnetic knots would be unavoidable, flattening the turbulence spectrum compared to the theoretical predictions. The latter contradicts both  Solar wind measurements and numerical calculations\footnote{If magnetic
reconnection is slow then, as was claimed by Don Cox (private communication), the interstellar medium should behave not like a fluid, but more like felt or Jello.}. 

Let us first consider the maximal rate allowed by the reconnection diffusion, i.e. evaluate the diffusivity arising from
the largest eddies. Dealing with reconnection diffusion one should consider all regimes of MHD turbulence, superAlfvenic, transAlfvenic and subAlfvenic. We start with a superAlfvenic regime, i.e. $M_A>1$. Magnetic field gets dynamically important as soon as its energy density exceeds the energy of eddies at  the Ohmic dissipation scale, which translates into Alfvenic velocity getting larger than the velocity of eddies at the Ohmic dissipation scale or the ion Larmor radius, whichever is larger. As the velocity in Kolmogorov turbulence scale as $v_l\sim l^{1/3}$, it is clear that even weak magnetic field can make a significant impact on the dynamics of the smallest eddies. In view of that it is advantageous to introduce a scale at which the magnetic field gets dynamically important and the nature of the turbulence changes from hydrodynamic to MHD (see Lazarian 2006), namely,
\begin{equation}
l_A=L(V_A/V_L)^3=LM_A^{-3}
\label{alf}
\end{equation}
where, similar to the notation in the hydrodynamic case, $L$ and $V_L$ are the injection scale and injection turbulent velocity, respectively. If the mean free path of
particles is larger than $l_A$, the scale $l_A$ may act as an effective mean free path in terms of particle diffusion along
magnetic fields. This is an important consideration for the diffusion of heat in collisionless fluid, but it is not so
for the diffusion of magnetic field and plasmas induced by turbulence. The corresponding diffusion coefficient for the maximal rate coincides with its hydrodynamic
counterpart given by Eq. (\ref{hydro}), i.e.
\begin{equation}
\kappa_{supA}=\kappa_{hydro}.
\label{supA}
\end{equation}
 This is due to the fact that for the largest eddies
of superAlfvenic turbulence are marginally affected by magnetic field. 

For low Alfvenic Mach numbers, i.e. for $V_A\gg V_L$ at large scales $\sim L$ the turbulence is weak (see Ng \& Bhattachargee 1997, LV99, Gaultier et al. 2000) and magnetic fields are slightly perturbed by propagating Alfven waves. The wave packets in weak turbulence evolve changing their
perpendicular scale $l_{\bot}$, while their scale $l_{\|}$ along the magnetic field does not change. The diffusion that is being induced by weak turbulence is substantially reduced compared to the case of hydrodynamic turbulence. It can be estimated as 
\begin{equation}
\kappa_{weak} \sim d^2\omega,
\label{kappa}
\end{equation}
 where $d$ is the random walk of the  field line over the wave period $\sim \omega^{-1}$. The weak turbulence at scale $L$ 
evolves over the non-linear evolution time (see review by Cho, Lazarian \& Vishniac 2003 and references therein)
\begin{equation}
\tau\sim (V_A/V_L)^2 \omega^{-1}
\label{tau}
\end{equation}
The transverse displacement of magnetic field lines in the perpendicular direction is a result of random walk
\begin{equation}
\langle y^2\rangle\sim (\tau \omega)d^2.
\label{y1}
\end{equation} 
According to LV99, the transverse displacement of magnetic field lines over distance $x$ is a spatial random walk given by the equation
$d\langle y^2\rangle/dx\sim L (V_L/V_A)^4$, which results in
\begin{equation}
\langle y^2 \rangle\sim L x (V_L/V_A)^4.
\label{y2}
\end{equation}
Combining Eqs. (\ref{kappa}), (\ref{tau}), (\ref{y1}) and (\ref{y2}) one gets the diffusion coefficient for the weak turbulence
\begin{equation}
\kappa_{weak}\sim LV_L (V_L/V_A)^3\equiv LV_L M_A^3
\label{kappa2}
\end{equation}
which is smaller than its hydrodynamic counterpart by the factor $M_A^3\ll 1$. 

The additional contribution to diffusivity in the case  $V_A\gg V_L$ comes from scales at which magnetic turbulence gets strong. As scaling of weak turbulence predicts $V_l\sim V_L(l_\bot/L)^{1/2}$ (LV99), at the scale 
\begin{equation}
l_{trans}\sim L(V_L/V_A)^2\equiv LM_A^2
\label{trans}
\end{equation}
the critical balance condition $l_{\|}/V_A\approx l_{\bot}/V_l$ is getting satisfied making turbulence strong. It is easy to see that the velocity corresponding to $l_{trans}$ is $V_{trans}\sim V_L (V_L/V_A)$. For strong turbulence the diffusion 
\begin{equation}
\kappa_{strong}\sim V_{trans} l_{trans}\sim L V_L (V_L/V_A)^3
\label{kappa_st}
\end{equation}
which coincides with Eq. (\ref{kappa2}), indicating that the enhanced diffusivity of smaller eddies in strong MHD turbulence regime can produce as efficient magnetic field mixing as the diffusivity induced by weak turbulence at the injection scale.
This coincidence illustrate the deep connection of the weak and strong turbulence in terms of the mixing that these
processes induce.

Dealing with subAlfvenic turbulence one should distinguish between the diffusivity parallel and perpendicular
to magnetic field. Magnetic field in a turbulent fluid changes the diffusion of plasma with particles moving
along magnetic field lines. At the same time, turbulent eddies in the direction perpendicular to the local direction of magnetic field, according LV99, are similar to the Kolmogorov picture. Indeed, in the presence of fast reconnection magnetic mixing is not inhibited for motions perpendicular to the local direction of magnetic field. As the local direction of the magnetic field varies in respect to the {\it mean} magnetic field, the diffusivity induced by mildly subAlfvenic and trans-Alfvenic turbulence is not expected to be very different for the directions parallel and perpendicular to the  {\it mean} magnetic field. At the same time, superAlfvenic turbulence, naturally, induces an isotropic diffusion.

The transAlfvenic case of $M_A\sim 1$ is the case which the GS95 model in its original formulation deals with. It should not
be special and should correspond to substituting $M_A\equiv 1$ in the expressions obtained for superAlfvenic and subAlfvenic cases, e.g.
in Eq. (\ref{kappa_st}).

As we mentioned earlier, the diffusion of magnetic field and plasmas at scales smaller than the scales of injection $L$ as well
as the scale transfer to the Alfvenic turbulence $l_{trans}$ for the case of subAlfvenic turbulence, is subject to Richardson diffusion (see
Eq. (\ref{Rich})). This is the type of diffusion that gets accelerated with time as well as with scale involved, i.e. "superdiffusion". We discuss below that dealing with star formation one is interested in the rate of the diffusion from a given clump or a cloud. In this case one can use the scale and
velocity dispersion of the cloud as the proxies of $L$ and $V_L$. In the case of subAlfvenic turbulence, the diffusion coefficient can be approximated as
\begin{equation}
\kappa_{cloud}\sim v_{cloud} l_{cloud} (v_{cloud}/V_A)^3
\label{kappa_cl}
\end{equation}
while for the case of turbulence being transAlfvenic or superAlfvenic the same estimate but without $M_A^3$ factor should
be used.

\subsection{Physical picture of reconnection diffusion in the absence of gravity}

The description above provides the mathematical framework of turbulent diffusivity in a homogeneous magnetized fluid. 
To what extent these results carry over to the mixing of highly inhomogeneous magnetic fields important for star formation is being discussed below. 

\begin{figure}
  \includegraphics[width=0.95\columnwidth,height=0.24\textheight]{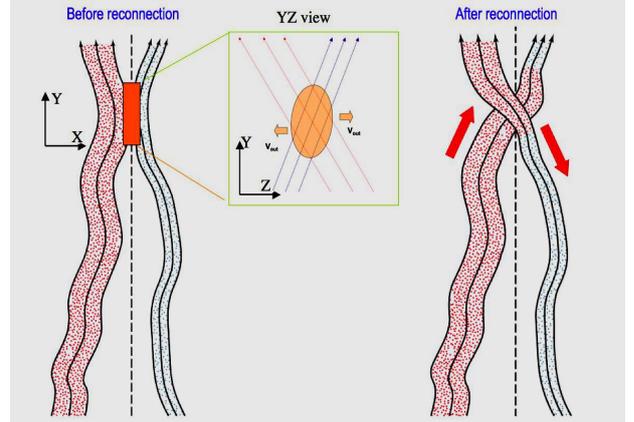}
  \caption{Motion of matter in the process of reconnection diffusion. 3D magnetic flux tubes get into contact and after reconnection plasma streams along magnetic field lines. {\it Left panel}: XY projection before reconnection, upper panel shows that the flux tubes are at angle in X-Z plane. {\it Right panel}: after reconnection. }
  \label{recdiff}
\end{figure}

We should mention that the concept of reconnection diffusion is based on LV99 model and was first discussed in Lazarian (2005) 
in the context of star formation. The concept, however, is also applicable to heat transfer (see Lazarian 2006) as well as the turbulent transfer of heavy elements, i.e. "metals", in the galaxy (see Lazarian 2011). The dynamo implications of the
LV99 reconnection model were discussed in LV99 and we may claim that the reconnection diffusion of dynamically important magnetic
fields should be used instead of the "turbulent diffusion" that was introduced to describe kinematic dynamo, i.e. the
dynamo applicable to the case of extremely weak magnetic fields only\footnote{It is important to stress that fast reconnection
does not change the helicity of the magnetic flux and therefore it does not solve the problem of helicity in astrophysical
dynamos (see Vishniac \& Cho 2001). Therefore the simulations where researchers use enhanced many orders of magnitude
resistivity in an attempt to mimic effects of turbulence smoothing on magnetic field are in error. However, the transport
of magnetic flux and smoothing that does not change helicity are important ingredients of any mean field dynamo that reconnection
diffusion takes care of.} (see \S 5.1). In terms of star forming clouds, the most important type
of dynamo is the turbulent dynamo (see Cho et al. 2010 and ref. therein). Such a dynamo may be essential for the clouds in
the early Universe when the plasmas were only weakly magnetized. Turbulent dynamo can also play role in increasing magnetization
of superAlfvenic turbulent clouds. However, within the present paper we are mostly concerned with the issues of magnetic field
removal via reconnection diffusion.    

If
magnetic field lines preserve their identify the diffusion of charged particles perpendicular to magnetic field lines is very restricted and 
the mass loading of magnetic field lines does not change. However, LV99 model (see also ELV11) suggests that the standard assumptions are violated if magnetized fluids are turbulent. 
As a result, in the presence of MHD turbulence, relative motions of plasma perpendicular
to magnetic field not only possible, but {\it inevitable}. 

We shall first illustrate the reconnection diffusion process showing how it allows plasma to move perpendicular to the mean inhomogeneous magnetic field (see Figure \ref{recdiff}). 
The set up is relevant to what we encounter in star formation. Two magnetic flux tubes with entrained plasmas intersect each other at an angle and due to reconnection the identity of magnetic field lines change. Before the reconnection plasma pressure $P_{plasma}$ in the tubes is different, but the total pressure $P_{plasma}+P_{magn}$ is the same for two tubes. This is a textbook situation
of a stable equilibrium. If plasmas are partially ionized then slow diffusion of neutrals which do not feel magnetic field directly can make gradually smoothen the magnetic field pressure gradients. We claim that in the presence of turbulence a different process takes place.

 Magnetic field lines in the presence of turbulence are not parallel. Such field lines can reconnect and do reconnect all the time in a turbulent flow (see \S 4.1). The process of reconnection changes the topology of the initial magnetic configuration and connects magnetic fields with different
 mass loading and plasma pressures. As a result, plasmas stream along newly formed magnetic field lines to equalize the pressure along flux tubes. Portions of magnetic flux tubes with higher magnetic pressure expand as plasma pressure increases due
 to the flow of plasma along magnetic field lines. The entropy of the system increases with magnetic and plasma pressures
 becoming equal through the volume. In other words, an efficient process of the diffusion of plasmas and magnetic field takes place and this process does
 not rely on the partial ionization of the material. In the absence of gravity, the effect of this process is to make magnetic field and plasmas more homogeneously distributed. The effective diffusion of both magnetic field and plasmas is about $LV_{stream}$, where $V_{stream}$ arises due to plasma pressure difference along the parts of the reconnected flux tube. Therefore it is of the order of the sound speed. A "microscopic" picture of the same process is presented in \S 6.  

If the process shown in Figure \ref{recdiff} were the only one, the speed of reconnection diffusion would be limited by the speed of plasma  motion along magnetic field lines. This is not the case, however. The exchange,
as a result of reconnection, of parts of flux tubes with entrained material is another process essential for
the process of reconnection diffusion. This ensures, for instance, that for supersonic turbulence, the exchange is happening at turbulent velocities.

\begin{figure}
\includegraphics[width=0.95\columnwidth,height=0.24\textheight]{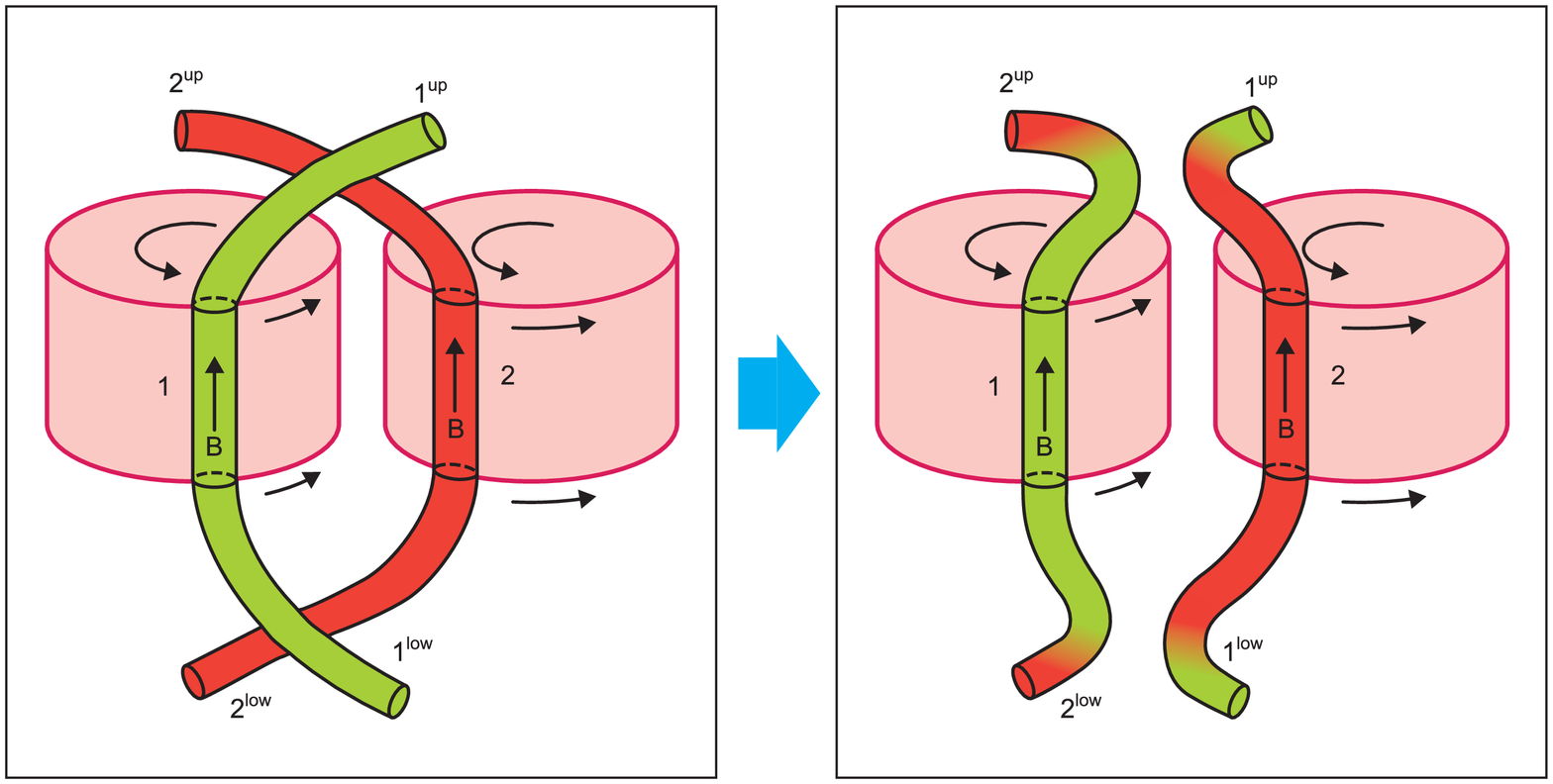}
\caption{Reconnection diffusion: exchange of flux with entrained matter. Illustration of the mixing of matter and
magnetic fields due to reconnection as two flux tubes of different eddies interact. Only one scale of turbulent
motions is shown. In real turbulent cascade such interactions proceed at every scale of turbulent motions.}
\label{mix}
\end{figure} 

To get a clear mental picture of what
is going on, consider a toy model of turbulence with only one scale of motions. Thus, similar to Figure \ref{recdiff}, where instead of considering the entire multitude of flux tubes of
different sizes we focused on what is happening with two flux tubes, here we consider just two neighboring eddies as is illustrated by Figure \ref{mix}. Magnetic flux tubes moving 
with the eddies reconnect and plasmas and magnetic fields associated with one eddy becomes a part of the other. This induces efficient diffusion of both magnetic field and plasmas.
Indeed, if, as in the case illustrated by Figure \ref{recdiff}, the densities of plasma along magnetic field lines are different in the two flux tubes, the reconnection in Figure \ref{mix}
creates  new flux tube with columns of entrained dense and rarefied plasmas. The situation is similar to the earlier discussed case with plasma moving along magnetic fields and equalizing the
pressure within the newly formed flux tubes. As a result, eddies with initially different plasma pressure exchange matter and equalize plasma pressure. This process can be described
as the diffusion of plasma perpendicular to the mean magnetic field. 

In reality, for turbulence with the extended inertial range, the shredding of the columns of plasmas with
different density proceeds at all turbulence scales making speed of plasma motion irrelevant for the diffusion. For the case of strong turbulence the diffusion of matter and magnetic
field is given by Eq. (\ref{kappa_st}).

Naturally, the process of reconnection diffusion also takes place when the pressure of plasmas is
the same throughout the volume. This is important for the diffusion of plasma impurities and heat.
 The mixing is happening as new magnetic flux tubes are constantly formed from magnetic flux tubes that belong to different eddies. It is clear that 
plasmas which were originally entrained on different flux tubes get into contact along the flux tubes created through reconnection. The process similar to the depicted one takes place at different scales down
to the scale of the smallest eddies. At the smallest scales, the microscopic diffusivity of plasmas takes over.

\begin{figure}
\centering
  \includegraphics[width=0.95\columnwidth,height=0.24\textheight]{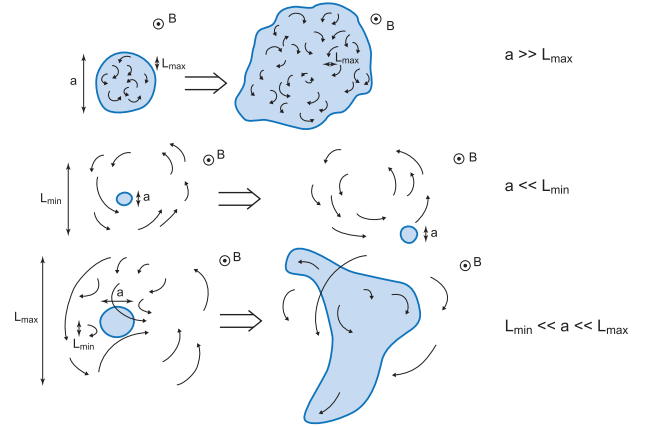}
  \caption{Reconnection diffusion depends on the size $a$ of the zone from which the diffusion happens. Different regimes emerge depending on the relation $a$ to the sizes of maximal and minimal eddies present in the turbulence cascade. Eddies perpendicular to magnetic field lines correspond to Alfvenic turbulence. The plots illustrate heat diffusion for different regimes. {\it Upper} plot corresponds to $a$ being less than the minimal size of turbulent eddies; {\it Middle plot} corresponds to $a$ being less than the damping scale of turbulence;  {\it Lower plot} corresponds to $a$ within the inertial range of turbulent motions. This is the case of Richardson diffusion.}
  \label{turb}
\end{figure}

The efficiency of reconnection diffusion depends on the scale of the motions involved. The process of reconnection diffusion may be illustrated with the case of 
the diffusion of impurity from a blob of the size $a$ (Figure \ref{turb}). This setting allows us to consider homogeneous turbulence, which is simplifies the analysis. 
 Turbulence is characterized by its injection scale $L_{max}$, its dissipation scale $L_{min}$ and its inertial range $[L_{min}, L_{max}]$. Consider Alfvenic eddies perpendicular to magnetic field lines. If turbulent eddies are much smaller than
$a$, i.e. $a\gg L_{min}$ they extend the spot acting in a random walk fashion. For eddies much larger than the blob, i.e. $a\ll L_{min}$  they mostly advect the spot\footnote{There is
also expansion of the spot arising from the Lyapunov deviation of the flow lines as we discuss in \S 6.1.}. If $a$ is the within the inertial range of turbulent motions, i.e. $L_{min}<a<L_{max}$ then a more complex dynamics of turbulent motions is involved. This is the case of Richardson diffusion (see
Eq. (\ref{Rich})).

If the blob $a$ is not just an impurity, but has density different from the density of the surrounding flow, the process of reconnection diffusion depends on the properties of turbulence within and outside the blob. The level and scale of turbulence may differ within and outside the blob. We may use the idealized picture in Figure \ref{turb} to get
a qualitative insight, nevertheless. If the blob is constrained by gravity, as this is relevant to star formation, and is turbulent up to the largest scale the diffusivity of magnetic field from the blob can be roughly estimated as $a v_a$, where $v_a$ is the velocity at the scale $a$. This estimate assumes that scale $a$ is within the range of strong MHD turbulence. Then the reconnection diffusion from the volume scales as $a^{4/3}$. If $a$ is within the range of weak turbulence the reconnection diffusivity is given by Eq. (\ref{kappa_cl}).

Our example in Figure \ref{turb} illustrates the diffusion perpendicular to magnetic field. As we mentioned in \S 3.2, Alfvenic motions that are most efficient in mixing perpendicular to the {\it local} direction of magnetic field. This direction, in general, does not coincide with the mean magnetic field direction. Therefore in the system of reference related to the mean magnetic field (and to the external observer) the diffusion of magnetic field and plasmas will happen both parallel and perpendicular to the mean magnetic field direction. However, the weaker the perturbations of the magnetic field, i.e. the smaller the $V_L/V_A$ ratio, the more anisotropic the reconnection diffusion is. 

\subsection{Reconnection diffusion in the presence of gravity}

In the 
presence of gravitational forces acting upon plasmas, the diffusion will also increase the entropy of the system, allowing
the "heavy fluid", i.e. gas, to be concentrated and "light fluid", i.e. magnetic field, to leave the the gravitational potential\footnote{The difference of the reconnection diffusion and ambipolar diffusion is that the former is associated with turbulent 
motions of matter.
Therefore, too intensive turbulence may upset the virial balance and instead of magnetic field diffusion, a dispersion of
the entire cloud can take place.}. 

Consider the idealized system of plasmas and magnetic field in a uniform directed downwards 
gravitational field with the acceleration $g$. In the
thermodynamic equilibrium the plasmas will have the Boltzmann distribution with $\rho \exp[-m_i g z/kT_{eff}]$, where
the effective temperature for supersonic turbulence can be roughly estimated from $kT_{eff}\sim m_{dom}V_L^2/2$, $m_{dom}$ is
the mass of dominant species in the flow. As a result, the light fluid, namely magnetic field, would
fill the entire volume and have the same pressure. Therefore the magnetization of the media measured in terms
of magnetic flux to plasma mass ratio is the lowest at the bottom and highest at the top of the system.

The redistribution of the magnetic field is induced by the reconnection diffusion. However,
one should clarify what is the diffusion coefficient for the magnetic field that is induced by the process. We discussed in
\S 5.2 that the diffusion happens in a different way at different scales. It is generally accepted that the turbulence
in molecular clouds is a part of a big power law turbulent cascade (see Figure\ref{CL} and the discussion in \S 3.1). The
gravitational cores that get rid of the magnetic field excess are much smaller than the scale of interstellar turbulent driving, i.e. $\sim 100$~pc. Assuming
that clouds are parts of the same turbulent cascade existing in the diffuse gas one can use the idealized sketch in Figure~\ref{turb} to estimate the efficiency of reconnection diffusion. It seems natural to associate the size of blob $a$
with the virial radius of the core $r_{vir}$, which for the case of the turbulence support of the core
is $G M/r_{vir}\sim V_{turb}^2$, where $G$ is the gravitational constant, $M$ is the core mass and $V_{turb}$ is the
turbulent velocity associated with the cloud at the scale $r_{vir}$.

\section{Reconnection diffusion and the identity of magnetic field lines}

Below we consider the process of reconnection diffusion microscopically, at the level of individual field lines.

\subsection{Explosive diffusion of magnetic field lines in turbulent flows}

A textbook description of magnetic field lines in perfectly or nearly perfectly conducting fluid assumes that the
line preserves its identity. There is a number of ways to see that this is impossible in a turbulent fluid (see ELV11).

We shall start by showing that the Richarson diffusion (see Eq. (\ref{Rich})) produces very non-trivial results. Consider the problem of separating particles in Kolmorogov turbulence. The separation between two particles $d/dt l(t)\sim v(l)\sim \alpha l^{1/3}$, where $\alpha$ is proportional to a cubical root of the energy cascading rate. The solution of this equation is 
 $l(t)=[l_0^{2/3}+\alpha (t-t_0)]^{1/3}$, which provides Richardson diffusion\footnote{Richardson diffusion presents an example
of {\it superdiffusion}, i.e. diffusion process for which $l^2\sim t^\beta$, $\beta>1$. The important consequences of Richardson
diffusion have been studied
for heat transfer and cosmic ray propagation (see Lazarian 2006, Yan \& Lazarian 2008).}  of $l^2\sim t^3$. However, as correctly
 stressed by Greg Eyink (2011), the odd feature of this solution is that the provides this type of
 fast separation even if the initial separation of particles is zero, which means the violation of Laplacian determinism. 
 Mathematically the above paradox is resolved by accounting to the fact that turbulent field is not differentiable\footnote{The Kolmogorov velocity field is Holder continuous, i.e. $|v(r_1)-v(r_2)|\leq C |r_1-r_2|^{1/3}$.}
 and therefore the initial value problem does not have a unique solution. Thinking physically, we cannot assume that
 the turbulence is present up to $l_0=0$. 

Although the previous example dealt with hydrodynamic turbulence, the essential features of this example cary over to the case of MHD turbulence, as in the plane perpendicular to the local direction of magnetic field, strong MHD turbulence satisfies the Kolmogorov description\footnote{This is not only similarity in terms of the spectrum. Cho, Lazarian \& Vishniac (2003) showed that in the local system of reference the intermittency of turbulence is also similar to the hydrodynamic one.}
 
\begin{figure}
\centering
  \includegraphics[height=.30\textheight]{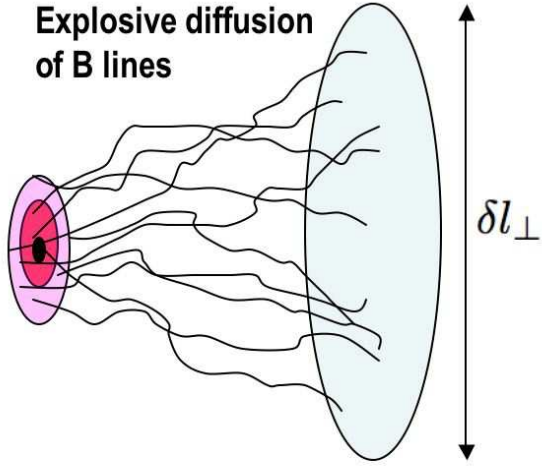}
  \caption{Particle tracing magnetic field lines may start at different initial locations shown as coaxial ellipsoids.
  However after a period of time the field line spread over a larger volume and the final position of the field lines
  does not correlate with their initial position.}
 \label{expl}
\end{figure}

 The study in LV99 revealed the Richardson-type diffusion of magnetic field lines. Figure \ref{expl} illustrates the loss of the Laplacian determinism for magnetic field lines. In analogy with the illustrative example above, the final line spread $l_{\bot}$ does not depend on the initial separation of the field lines. This is a remarkable effect that provides a microscopic picture of reconnection diffusion based on the description of magnetic field lines rather than on the reconnection of well organized flux tubes in \S 5.1.
 
\begin{figure}
\centering
  \includegraphics[height=.40\textheight]{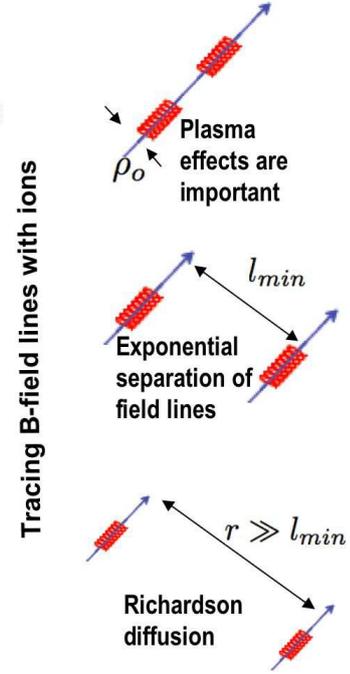}
  \caption{{\it Upper plot} Ions tracing the same magnetic field line. The diffusion and decorrelation arises
  from plasma or Ohmic effects as well Rechester-Rosenbluth effect. {\it Middle plot} Ions separated by scales 
  much larger than the ion Larmor radius are further separated by the Rechester-Rosenbluth effect. {\it Lower} At scales larger than the turbulence damping scale the Richardson diffusion takes over resulting in explosive
  separation of field lines.}
 \label{regimes}
\end{figure}

 We shall consider tracing magnetic field lines in the realistic turbulence with the dissipation scale $l_{min, \bot}$,
 where, as everywhere in the paper, $\bot$ denotes the scale perpendicular to the local magnetic field. We feel that this avoids some of the paradoxes discussed in ELV11 and also allows to treat a more generic case of astrophysical turbulence. However, we first briefly stress a couple of points presented in ELV11.
 First of all, resistivity, whatever its nature, introduces stochastic forcing in the description of magnetic field line dynamics. Indeed, the induction equation with the resistive term $\eta\Delta B$ signifies stochasticity associated with Ohmic diffusion. Therefore the definition of the magnetic field line on scales affected by resistivity is not deterministic. In addition,
 one has to accept that the magnetic field line motion is a concept defined by convention and not testable experimentally. This point is stressed in the literature (see Newcomb 1958,
 Vasylianas 1972, Alfven 1976, ELV11), but sometimes forgotten. Thus magnetic field lines may be tagged by ions that start at the same field line (see Figure \ref{regimes}) In the case of smooth laminar magnetic field and ideal MHD equations the motions of ions will reveal magnetic field lines and two ions on the same field line will always remain on the same line. The situation is radically different in the presence of turbulence.

In our thought experiment we shall trace ions moving with the same velocity and
separated perpendicular to magnetic field by a distance of a Larmor radius $\rho_0$. As we discuss further, if the separation is
less $rho_0$ than that one can appeal to plasma effects to increase the separation to $\rho_0$. Let us assume that 
the minimal scale of turbulence $l_{min, \bot} >\rho_0$. In this situation the dynamics of ions
 can be approximated by the dynamics of charged particles in toy model of "a single scale MHD turbulence" discussed in Rechester \& Rosenbluth 
 (1978). Indeed, the turbulent motions at the critically damped scale $l_{min, \bot}$ are dominant for shearing and steering matter and magnetic field on the smaller scales. The Rechester \& Rosenbluth (1978) theory predicts the Lypunov growth 
of the perpendicular separation of
 ions, i.e. the separation gets $\rho_0\exp(l/l_{min, \|}$, where $l$ is the distance traveled by ions and $l_{min, \|}$
 is the scale parallel scale of the critically damped eddies with the perpendicular scale $l_{min, \bot}$ (see
 also Narayan \& Medvedev 2001, Lazarian 2006). Thus to get separated by the distance $l_{min, \bot}$ the ions 
 should travel the so-called Rechester-Rosenbluth distance:
 \begin{equation}
 L_{RR}\approx l_{min, \|} ln(l_{min, \bot}/\rho_0)
 \label{RR}
 \end{equation}
 which at most a dozen times larger than the microscopic scale $l_{min, \|}$. As soon as ions get separated over
 the distance of $l_{min, \bot}$ they get into different eddies and the process of Richardson diffusion starts. Thus
 after a relatively short period when ions move in a correlated manner remembering their original position, a
 stochastic regime when the initial vicinity of the ions is completely forgotten takes over. As
 we used the ions as tracers of magnetic field, we can talk about the stochasticity of magnetic field lines (as traced by
ions).

 If the minimal turbulence scale is equal to $\rho_0$, the arguments above only get stronger, as from the very beginning
 the ions may experience stochastic turbulence driving and get uncorrelated. In many cases, for instance, in fully or
 mostly ionized ISM (e.g. with the ionization larger than 93\% as discussed e.g. in Lazarian, Vishniac \& Cho 2004),
 the Alfvenic turbulent cascade at $\rho_0$ gets continued as a whistler cascade involving only electrons. Such
 cascading provides stochasticity below $\rho_0$. With the whistler scaling as $v_l\sim l^{2/3}$ (see
 Cho \& Lazarian 2009 and references therein) the "whistler-induced" Richardson diffusion should go as 
 $l^2\sim t^6$, inducing fast separation
 of magnetic field lines that can be now traced by electrons and inducing stochastic perturbations on ion trajectories.
 This only makes the case for the stochasticity when the ions are separated initially over less than 
 $\rho_0$ more evident.
 
 Consider the case of ions separate by less than the $\rho_0$ distance and turbulence truncated at scales larger
 than $\rho_0$. The generalized Ohms law can be written as 
 \begin{eqnarray} 
&&{\bf E}=-\frac{1}{c}{\bf u\times B}+\eta_\perp{\bf J}_\perp+\eta_\|{\bf J}_\| 
+\frac{\bf J\times B}{nec}\cr
&&   \,\,\,\,\,-\frac{{\bf \nabla \dot P_e}}{nec} 
+ \frac{m_e}{ne^2}\left(\frac{\partial {\bf J}}{\partial t} + \nabla\dot({\bf uJ}+{\bf Ju}-\frac{1}{ne}{\bf JJ})\right), 
\label{Ohm} 
\end{eqnarray}         
when quasi-neutrality and $m_e\ll m_i$ are assumed (see Bhattacharjee et al. 1999). 
 The electric fields on the righthand 
side are, respectively, the motional field, Ohmic fields associated to perpendicular and 
parallel resistivities, the Hall field, a contribution from the electron pressure tensor, and 
electron inertial contributions.  All of these terms 
contribute to the slippage of magnetic field-lines through the plasma on the scales of the order of $\rho_0$. 
 This introduces small scale diffusion and therefore
 stochasticity for the ions initially very close in their position (ELV11). Therefore the ions are bound to separate fast
 however close they were initially. The diffusion introduced by the terms in the Ohm's law is important on the scales
 $\rho_0$, but for the reconnection diffusion the exact form of these terms does not matter. The initial separation
of the particles is being fast forgotten and it is the large scale turbulence that determines the macroscopic stochasticity of magnetic field lines.

For the purposes of discussing reconnection diffusion, we have chosen ions as our trace particles. One, however, may argue
 that electrons as current carrying agents are more appropriate. This does not change our arguments above, however,
 as using the Larmor radius of electron $\rho_{electron}$ instead of $\rho_0$ in Eq. (\ref{RR}) changes the result by
 an insignificant factor.

 \subsection{Spontaneous stochasticity of magnetic field lines and reconnection diffusion}
 
 The process of magnetic field lines becoming stochastic in turbulent fluids is also called "spontaneous stochasticity" (see Eyink 2011, ELV11). We believe that the way of tracing magnetic field lines with ions discussed in \S 6.1 has a very
 transparent meaning in terms of diffusion of plasmas in astrophysical conditions. The common wisdom underlying
 the star formation research, for instance, is based on the picture of laminar field lines where nearby ions stay all the time
entrained on the same field line provided that the non-ideal effects, e.g. resistivity, are negligibly small. It is evident from
 \S 6.1 that this is not correct for realistically turbulent fluids. 

\begin{figure}
\centering
  \includegraphics[height=.22\textheight]{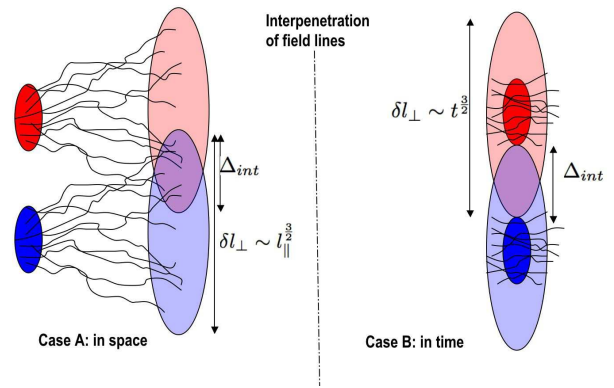}
  \caption{Microscopic physical picture of reconnection diffusion. Magnetized plasma from two regions is spread
  by turbulence and mixed up over $\Delta_{int}$. {\it Left panel}: Description of the process in terms of field wandering in space. {\it Right panel}. Description of the spread in time.}
 \label{alt}
\end{figure}

To proceed with our discussion of the physics of reconnection diffusion, consider two turbulent volumes at a distance between them (see Figure \ref{alt}). Each of the volumes has its
own set of magnetic field lines. However, as the field lines wander and spread as the consequence of the Richardson diffusion they overlap in the volume $\Delta_{int}$ and their identity as associated
with the particular volume is lost (Figure \ref{alt}, left). Magnetic field lines reconnect and the newly formed lines allow plasma exchange between the initially disconnected volumes. If plasma and magnetic field pressures in the volumes was different, one can easily seen that this picture based on the diffusion of magnetic field lines is
analogous to the picture describing the exchange of plasmas between magnetic flux tubes in Figure \ref{recdiff}.

Figure \ref{alt}, left, illustrates the spread of magnetic field lines in the perpendicular direction as magnetic field lines are traced by particles
moving along them. Using Eq. (\ref{y2}) one can get the RMS separation of the magnetic field lines (LV99)
\begin{equation}
\delta l_{\bot}^2\approx \frac{l_{\|}^3}{L}\left(\frac{V_L}{V_A}\right)^4,
\label{l_rms}
\end{equation}
i.e. $l_{\bot}^2$ is proportional to $l_{\|}^3$. This regime identified in LV99 is a Richardson diffusion
but in terms of magnetic field lines\footnote{This regime induces perpendicular "superdiffusion" in terms of cosmic rays and other charged
particles streaming along magnetic field lines.}. The numerical testing of this prediction of LV99 is shown in Figure \ref{wand}. This is the case
of diffusion in space. Eq. (\ref{l_rms}) allow one to calculate the distance $l_{\|}$ at which the magnetic field lines of regions
separated by $l_{\bot}$ start overlapping\footnote{Incidentally, in terms of reconnection, Eq. (\ref{l_bot}) expresses the thickness of outflow 
denoted as $\Delta$ in Figure \ref{LV}. Substituting this value in Eq. (\ref{vrec}) one recovers the LV99 reconnection rate, i.e. Eq. (\ref{LV99}).},
i.e. $l_{\|}\approx (\delta l_{\bot}^2 L)^{1/3}(V_A/V_L)^{4/3}$. Thus for sufficiently large $l_{\|}$ all parts of the volume of magnetized plasmas
get connected. In other words, the entire volume becomes accessible to particles moving along magnetic field lines. Naturally, in this situation
the customary for the star formation community notion of flux to mass ratio loses its original sense.   

In addition, in turbulent plasmas, our turbulent volumes spread due to Richardson diffusion with $\delta l_{\perp}^2\sim t^3$. This process is
illustrated by Figure \ref{alt}, right. The magnetized plasmas spreads by subAlfvenic turbulence over a larger volume, while motions of plasma along magnetic field lines are ignored. The two initially disconnected volumes get overlapped and the magnetic field and plasma of two regions gets mixed up. Again, one easily can see that the process that we described in terms of magnetic field lines is similar to the one we described in terms of reconnected magnetic field flux tubes 
in Figure \ref{mix}. This is expected, as magnetic reconnection is an intrinsic part of the picture of MHD turbulence that
governs the dynamics of magnetic field lines (LV99)\footnote{One can also claim that spontaneous stochasticity of magnetic field in turbulent fluids is an underlying process that governs magnetic reconnection (ELV11). As
we mentioned earlier, fast reconnection makes MHD turbulence theory self-consistent.}

For the partially ionized gas the following questions are practically important. Does the Rechester-Rosenbluth length given
by Eq. (\ref{RR}) present a serious bottleneck for Richarson diffusion on the larger scales? Does the Richarson diffusion
provide faster diffusion rates compared to the ambipolar diffusion? The latter question is addressed in \S 11.4 where we argue
that if ambipolar diffusion is too efficient, we expect turbulence to decay, while if strong MHD turbulence is present in the 
volume this means that the reconnection diffusion dominates. The answer to the former question is more subtle. There
has not yet been a study of the Richardson diffusion in a partially ionized gas. However, appealing to the equivalence
of fast reconnection and spontaneous stochasticity revealed in ELV11, we can appeal to the work by Lazarian et al. (2004) 
that demonstrates that the reconnection in a partially ionized gas is fast. This justifies our application of the Richardson diffusivity to the partially ionized gas. Intuitively, one can argue that the Richardson diffusion on the scales larger
than the damping scale makes the latter irrelevant similar to the case of damping scale in fully ionized plasma that we
discussed earlier\footnote{In fact, Lazarian et al. (2004) show that the turbulence in partially ionized gas demonstrates several regimes, including the intermittent "resurrection" of turbulence cascade at scales less that the ambipolar damping scale. These effects make magnetic fields stochastic on scales much less that the naively estimated $l_{min, \|}$ that enters Eq. (\ref{RR}).}.

Naturally, in the presence of diffusion, ions that we use to trace magnetic field spread in the volume, reproducing the
results of spreading of magnetic field in the turbulent volume via reconnection diffusion that we argued in \S 5.2. In the 
presence of gravity, it is obvious that the lighter fluid of magnetized ions should tend to escape the gravitational potential
as we discussed in \S 5.3. In other words, tracing of ions provides us with the microscopic picture of reconnection diffusion.

\subsection{Magnetic field wandering and reconnection diffusion}

\begin{figure}
\centering
  \includegraphics[height=.30\textheight]{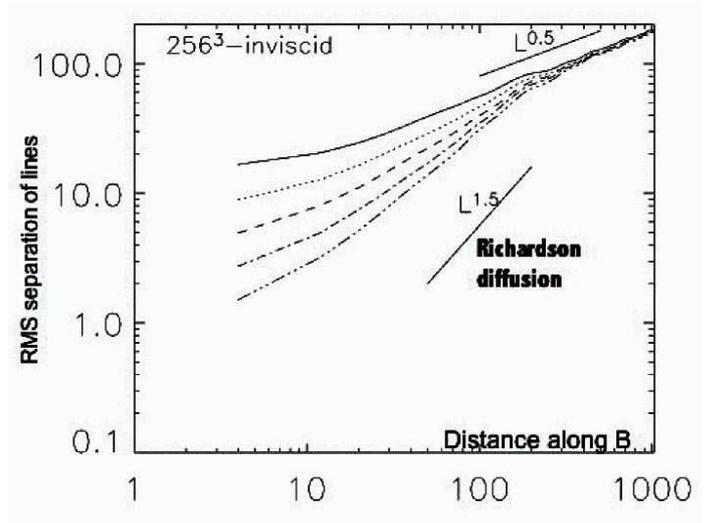}
  \caption{Magnetic field wandering from numerical simulations. Separation of magnetic field lines shown. Modified from Lazarian et al. 2004.}
 \label{wand}
\end{figure}

Magnetic field wandering described in LV99 is a implementation of Richardson diffusion 
(ELV11). This very wandering has then been then used in the literature to describe
 the properties of cosmic ray diffusion and heat transport (see Narayan \& Medvedev 2001, Lazarian 2005, Yan \& Lazarian 2008)
 and has been tested numerically (Maron et al. 2004, Lazarian et al. 2004). Figure \ref{wand} shows 
 numerical results on magnetic field separation. The regime corresponding to the Richardson-type scaling
 is clearly seen at the scales corresponding to the inertial range of the turbulence. At the scales larger than
 the injection scale the usual diffusion regime takes over. The latter regime was discussed decades ago 
within earlier models of MHD turbulence (see, for instance, classical papers
 by Parker 1965, Jokipii 1973).

Neither of the regimes of magnetic wandering can 
be understood as static with magnetic field lines preserving their identify in turbulent flows. Magnetic field lines
should constantly reconnect inducing the exchange of plasmas and magnetic field inducing the reconnection diffusion 
process that we advocate in this paper. In view of this, interesting ideas on the non-trivial nature of magnetic
field lines can be found in a prophetic book by Parker (1979). Nevertheless, Parker (1979) does not formulate the
reconnection diffusion or spontaneous stochasticity concepts. 

\subsection{MHD and plasma-based descriptions of reconnection diffusion}

In the previous section we proved that LV99 ensures efficient diffusion of matter and magnetic fields in turbulent fluids. One
may wonder whether other models of fast reconnection, e.g. collisionless reconnection, can also induce reconnection diffusion.

First of all, we would like to stress that the LV99 model is not in conflict with the studies of magnetic reconnection in collisionless plasmas that have been a major thrust of the plasma physics community (see Shay et al. 1998, Daughton et al. 2006). Unlike latter studies, LV99 deals with turbulent environments. It shows, as we discussed in \S 4.1 that local reconnection rates are influenced by plasma effects, e.g. kinetic effects of Hall effects, but the overall or global reconnection rate, i.e. the rate at which magnetic flux tubes reconnect, is determined by the turbulent broadening of the reconnection region. Thus, in the turbulent astrophysical media the rate of reconnection is not going to be affected by additional mechanisms. As a result, the reconnection diffusion will proceed with its maximal rate limited by turbulent motions only.

If, on the contrary, the media that we deal with is not turbulent, reconnection diffusion does not take place even in the presence
of fast reconnection. Turbulent mixing is a necessary condition for reconnection diffusion to exist. If other mechanisms of
reconnection induce turbulence, this turbulence will induce reconnection diffusion and we return back to the case above.

Plasma effect might be potentially important on a more subtle level, however. As field wandering is essential for reconnection diffusion, one may wonder to what extend the plasma effects can be neglected while describing field wandering. 
Indeed, as we mentioned in \S 6.1 the non-ideal terms in the Ohms law (\ref{Ohm}) provide the stochasticity of charge carriers and
therefore of the magnetic field lines that these charges trace. Consider, for instance, Hall term which is most commonly 
invoked in the literature of fast reconnection. Hall term was also invoked in a studies of magnetic field loss by circumstellar accretion disks\footnote{To see the effect the authors had to adopt the Hall term much larger than its
value for the adopted parameters of the media.} (see Krasnopolsky et al. 2010). The usual criterion for the Hall term to dominate is that the electron flow velocity
is dominated by the current. However, to dominate magnetic field stochasticity the criterion should be different as the 
correlations of the Hall velocity are short ranged. Assuming the small-scale equipartition of velocity and magnetic field,
$B(r)^2/4\pi\sim \rho v(r)^2/2$ and turbulent correlation of velocities $\langle v_i v_j\rangle\sim C r^{\gamma}$, one
gets for the Hall velocity $V_{Hall}=J/ne=c\nabla \times B/4\pi ne$ correlations
\begin{equation}
\langle V_{Hall, i} V_{Hall, j}\rangle \sim \left(\frac{c}{4\pi ne}\right)^2 \Delta \langle B_i B_j\rangle
\label{Hall}
\end{equation}
The right hand sight Eq. (\ref{Hall}) can be estimated as $\left(\frac{c}{4\pi ne}\right)^2 4\pi n \rho C r^{\gamma -2}$, which is
much smaller than the correlation of the turbulent velocities if the distance between point of correlation 
$r\gg c^2 m_i/4\pi n e^2=\delta_i$, where $\delta_i$ is the ion inertial skin depth. This estimate is consistent with a more
elaborate one in ELV11. Therefore even large Hall velocities do not affect meandering of magnetic field lines on the scale
much larger that the inertial skin depth. The reconnection diffusion applicable to star formation deals with scales 
$\gg \delta_i$. 

To finish with the discussion of plasma effects, we should mention that the Hall MHD (HMHD) is frequently presented as a proper way to describe reconnection in astrophysical systems. However, it is shown in ELV11 that HMHD is rarely applicable to the actual astrophysical plasmas. Indeed, the derivation of Hall MHD based on collisionality requires that the ion skin-depth $\delta_i$ 
must satisfy the conditions $\delta_i\gg S \gg \ell_{mfp,i}$, where $S$ is the scale of large-scale variations of
magnetic field.  The second inequality is needed so that a two-fluid 
description is valid at the
scales of interest, while the first inequality is needed so that the Hall term remains 
significant at those scales.  However, substituting $\delta_i=\rho_i/\sqrt{\beta_i}$ into the expression for the Coulomb
collisional frequency (see Eq. (\ref{lmfp-rho})) yields the result
\begin{equation} 
\frac{\ell_{mfp,i}}{\delta_i}\propto \frac{\Lambda}{\ln\Lambda}\frac{v_{th,i}}{c}. 
\end{equation}
where $\Lambda=4\pi \rho n \lambda_D^3$ is the number of particles in the Debye sphere. For weakly 
coupled astrophysical plasmas $\Lambda$ is really large (see table in EVL11) and therefore 
$\ell_{mfp,i}\gg \delta_i,$ unless the ion temperature is extremely low.  Thus, Hall MHD is valid 
only for cold, dense plasmas,e.g. that produced by the MRX reconnection experiment (e.g. Yamada et al. 2010), but
not in the conditions of the diffuse ISM and molecular clouds where star formation takes place. 

\section{Hypothetical weak regime of reconnection diffusion}

Our discussion above shows that, similar to the case of hydrodynamic turbulence, strong MHD turbulence induces efficient diffusion. In this regimes the initial separation of particles and magnetic field lines does not affect their final position after a sufficiently extended period of time (see our illustration of the Richardson diffusion in \S 6.1). For superAlfvenic and transAlfvenic turbulence the regime of accelerated diffusion spans from the scale of dissipation to the one of injection. For subAlfvenic turbulence the Richardson regime
covers scales from the dissipation scale to the scale of the transition to the strong turbulence, i.e. $t_{trans}$ given
by Eq. (\ref{trans}). The latter is true provided that the dissipation scale is larger than $t_{trans}$. 

What would happen if the turbulence dissipation scale is larger than $t_{trans}$? This can result in a very different regime of diffusion. Indeed, the diffusion induced by weak turbulence obeys the ordinary diffusion laws with the squared displacement
proportional to time, i.e. $\delta^2\sim t$, rather than $t^3$ for the Richardson diffusion. The trajectories of particles exponentially
depart, but they remember their starting point. Therefore the reconnection is expected to depend on the plasma microphysics and
will not be fast. This is a very special regime of very weak driving and very strong dissipation.  This regime can be called "weak reconnection diffusion". In this situation, one would expect interesting new effects, e.g.  the joined action of ambipolar diffusion and weak reconnection diffusion. 

We feel that the situation with weak reconnection diffusion requires further studies and no definitive conclusions are possible at the moment. Potentially, weakly driven turbulence in a partially ionized gas could realize "weak reconnection diffusion". However, Lazarian et al. (2004) claimed that in the partially ionized gas the turbulence proceeds first as a viscosity damped magnetic turbulence with a shallow spectrum of magnetic perturbations and steep velocity spectrum. Then this regime transfers to strong turbulence involving only ions and electrons.
From the scale of the strong turbulence resurrection to the dissipation scale one expect to observe the regime of the accelerated
diffusion making the microphysics on the smaller scales irrelevant. A more fundamental problem is that in the regime of weak turbulence without a transition to the strong one, the reconnection may get dependent on resistivity, i.e. become slow. In this situation one would expect the accumulation of magnetic winding and also eventual transfer to bursty strong regime even in one
fluid approximation. These issues are, however, are beyond the scope of the present paper. For astrophysical settings that we discuss in \S 8 and \S 9 we
do not expect the reconnection diffusion to be in weak regime.

\section{Theoretical expectations and numerical simulations of reconnection diffusion}

\subsection{Limitations of numerical simulations}

Recently we performed a few numerical studies to explore the consequences of reconnection 
diffusion. Similarly, as numerical studies of ambipolar diffusion do not "prove" the very concept of
ambipolar diffusion, our studies were not intended to "prove" the idea of reconnection diffusion. 
Our goal was to demonstrate that, {\it in agreement with the theoretical expectations}, the process of
reconnection diffusion is important for a number of astrophysical set-ups relevant to star formation.

We have to admit that the limitations arising from numerical simulations are not always appreciated within the astrophysical
community. While low resolution observations provide a true picture smoothed by a telescope beam, potentially,
low resolution numerical simulations may provide a completely erroneous physical picture. 

To understand the difference between reconnection in astrophysical situations and 
in numerical simulations, one should recall that the dimensionless combination that
controls the resistive reconnection rate is the Lundquist number given by Eq. (\ref{Lun})\footnote{The magnetic
Reynolds number, which is the ratio of the magnetic field decay time to the eddy
turnover time, is defined using the injection velocity $v_l$ as a characteristic
speed instead of the Alfv\'en speed $V_A$, which is taken in the Lundquist
number.}. Because of the huge astrophysical 
length-scales $L_x$ involved, the astrophysical Lundquist numbers are also huge,
e.g. for the ISM they are about $10^{16}$, while present-day MHD simulations
correspond to $S<10^4$. As the numerical efforts scale as $L_x^4$, where $L_x$
is the size of the box, it is feasible neither at present nor in the foreseeable future to have simulations with 
realistically Lundquist numbers. Therefore our numerical studies of reconnection diffusion in 
Santos-Lima et al. (2010, 2012) deal with a different domain of Lundquist numbers and the theoretical justification
why {\it for the given problem} the difference in the Lundquist numbers is not essential is mandatory. For the case of
reconnection diffusion simulations, LV99 theory predicts that the dynamics of reconnection is independent from the Lundquist number and therefore 
the reconnection in the computer simulations {\it in the presence of turbulence} adequately represents the astrophysical 
process. 

\begin{figure}
\center
\includegraphics[width=0.9\columnwidth]{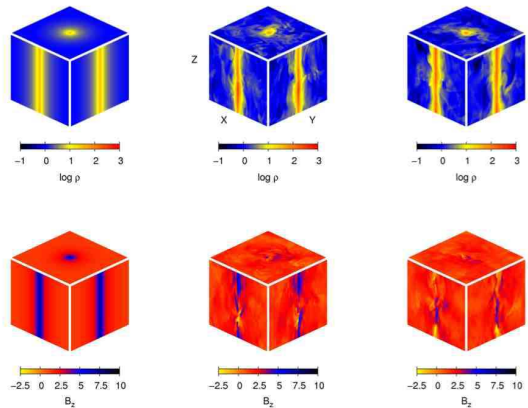}
\caption{Removal of magnetic field via reconnection diffusion from cylindrical models of molecular clouds. In the process of
simulations the density is accumulated at the center of the potential well ({\it upper raw}), while the magnetic field
leaves the center ({\it lower raw}). From Santos-Lima et al. (2010).
\label{10}}
\end{figure}

We can rely on LV99 theory as by now its {\it analytical
predictions} have been confirmed (see Kowal et al. 2009, 2012, Lazarian et al. 2011, Vishniac et al. 2012; see also \S 4.3). Moreover, the correspondence of the 
LV99 theory and more recent ideas on spontaneous stochasticity developed in the theory of turbulence have been revealed in ELV11. These studies
provide us with confidence that we understand what is happening with the magnetic field in the  Santos-Lima (2010, 2011) simulations. 

We must also stress that for the reconnection diffusion process turbulence is essential. Similarly with the 
Lundquist number issue, the perfect representation of astrophysical turbulence is not possible due to
the huge difference of the Reynolds numbers in astrophysical fluids and in simulations. However, our theoretical
study allows us to claim that for the reconnection diffusion the scales of turbulence injection or the scale of the transition to strong MHD
turbulence ($l_{trans}$, see Eq. (\ref{trans})) matter the most. Therefore if the simulations resolve these scales one may rely on the reconnection diffusion
in astrophysical settings being correctly represented. In terms of simulating reconnection diffusion, it is important to keep in mind that the LV99 model
predicts that the largest eddies are the most important for providing outflow in the reconnection zone and therefore the reconnection will not be appreciably changed if turbulence does not have an extended inertial range. In addition, LV99 predicts that the effects of anomalous resistivity, including that arising from the finite numerical grid, do not change the rate of turbulent reconnection (see more discussion in \S 4.2).

\subsection{Reconnection diffusion and magnetic field removal from clouds} 

Understanding of the nature of reconnection diffusion allows one to simulate process with 3D MHD codes. Some results 
of such a simulation is shown in Figure \ref{10}. We observe that {\it in the absence of ambipolar diffusion}, the magnetic
field escapes the gravitational potential, allowing the matter to become concentrated in the center. The simulations were
performed both from equilibrium configurations simulating subcritical
clouds and in collapsing clouds, simulating supercritical clouds. In all the cases, the efficient removal of magnetic flux from clouds
was observed. 

It worth stressing that the application of concept of reconnection diffusion is not limited to accounting for the known observational
facts. It includes predictions, e.g. the possibility of a gravitational collapse irrespectively of the degree of cloud ionization. 
As we discuss in \S 9.4 reconnection diffusion provides an attractive scenario for star formation in special cases when the expected high ionization should make ambipolar diffusion prohibitively slow.

\subsection{Reconnection diffusion and circumstellar accretion disks}

Circumstellar disks are known to play a fundamental role at the late stages of star formation (see Aikawa \& Nomura 2008). Observational
studies revealed that embedded magnetic fields in molecular cloud cores are high enough to
inhibit the formation of rotationally supported disks. Ambipolar diffusion is not powerful enough to induce
the removal of magnetic fields fast enough. This motivated Shu et al. (2006) to talk about effects of
enhanced resistivity that can explain the observational data (see also more recent elaborations of
this idea of microscopic resistivity in Krasnopolsky et al. 2010 and Li et al. 2011). On the contrary, appealing to fast
reconnection in LV99, Lazarian \& Vishniac (2009) argued that the removal of magnetic field is
due to reconnection diffusion. 

\begin{figure}
\center
\includegraphics[width=0.9\columnwidth]{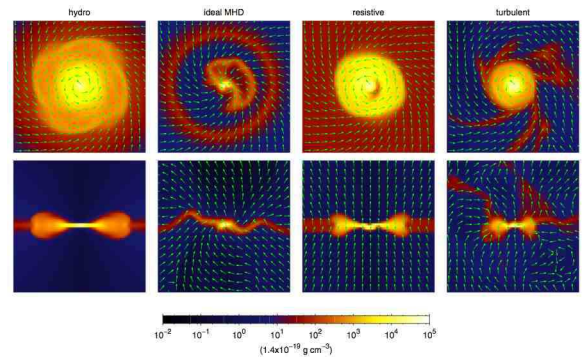}
\caption{Formation of circumstellar disks (from left to right): in hydro simulations, MHD simulations without turbulence,
MHD simulations with unrealistically high resistivity and MHD simulations with 
turbulence at the start of simulations. Reconnection diffusion produces realistic disks. From Santos-Lima et al. (2012).
\label{100}}
\end{figure}

Figure \ref{100} shows results of recent simulations by Santos-Lima et al. (2012) which, indeed, support the 
notion that reconnection diffusion is the process that is responsible for the removal of magnetic fields
from accretion disks. The turbulence injection in the simulations was done to mimic the turbulence
associated with the process of disk formation. The simulations also testify that without turbulence
for realistic parameters of resistivity the formation of disks is suppressed. This work established the correspondence of
the properties of the disks produced by reconnection diffusion to observations. While
additional processes may also be important for solving "magnetic braking catastrophe" (see Seifried et al. 2012), the reconnection diffusion is the process that is  definitely present\footnote{In comparison the models with enhanced
resistivity operate with the resistivities that are not motivated by the known physics.} in turbulent fluids and therefore is an essential
part of circumstellar accretion disks models. Apart from being important for the dynamics of accretion disks, the intensity of turbulent magnetic
field is important for the growth of dust particles in accretion disks (Steinacket et al. 2010) through the coagulation and
shuttering induced by MHD turbulence (see Lazarian \& Yan 2002, Yan \& Lazarian 2003, Hoang, Lazarian \& Schlickeiser 2012). Thus the process should be taken into account for any self-consistent model of circumstellar accretion disks (e.g. see Henning \& Meeus 2011), including accretion disks around massive stars (see Keto \& Zhang 2011).    

\section{Predictions and tests for reconnection diffusion}

Reconnection diffusion is a physical process very different from the ambipolar diffusion. 
Thus it is not surprising that the star formation controlled by reconnection diffusion is
very different from the picture based on the ambipolar diffusion concept.
Below we outline a few predictions that follow from the reconnection diffusion model.

\subsection{Reconnection diffusion in interstellar diffuse gas}

A naive picture of frozen in magnetized plasma suggests that fluctuations of magnetic field and density should
be correlated. The correlations may not be perfect, as motions along magnetic field lines that compress
only gas are also present, e.g. slow modes in the media with magnetic pressure larger than the
pressure of the ionized gas (see Cho \& Lazarian 2002, Passot \& Vazquez-Semadeni 2003). 

Observations by Troland \& Heiles (1986) demonstrated a poor correlation between the magnetic field 
strength and density, which was rather unexpected at the time the work was done. Note that the degree of ionization of
the diffuse media is sufficiently large to make the effects of ambipolar diffusion negligible. We discuss the approaches to
this problem based on the works by Passot \& Vazquez-Semadeni (2003) and Heitsch et al. (2004) in \S 11.6 and \S 11.4, respectively.

In view of our earlier discussion in this paper, the above result is expected. Indeed, we discussed that the reconnection diffusion tends to
make the magnetic energy density uniformly distributed in the volume. In other words,  in the presence of 
reconnection diffusion mixing of density fluctuations by turbulent eddies is present and this
should mitigate any density-magnetic field correlations arising from simultaneous compression of magnetic field
and conducting gas\footnote{For highly supersonic turbulence shock formation may somewhat alter the picture
and the correlation of magnetic field and density is being observed in supersonic simulations (Burkhart et al. 2009). 
However, the bulk of the turbulence in warm diffuse media for which the observations are applicable is 
subsonic (see Burkhart et al. 2012).}

Applying our results in \S 5.1 one can conclude that the diffusion of magnetic field for the diffuse interstellar media is determined by motions at the large scale which for super- or transAlfvenic driving provides the coefficient of
reconnection diffusivity $L V_L$ if the scale of interest $a$ is larger than the turbulence injection scale $L$. Similarly, for
$a<L$ the diffusion coefficient is $\sim a V_L (a/L)^{1/3}$ if $a<L$. For the subAlfvenic driving, i.e. for the 
Alfven Mach number of
turbulence $M_A<1$ additional factor of $M_A^3$ is expected in the diffusion coefficient expressions reflecting
the decreased efficiency of mixing by subAlfvenic turbulence. The results of numerical simulations are consistent with
these expectations (see Santos-Lima et al. 2010). In fact, reconnection diffusion was demonstrated to decorrelate magnetic
field and density, if initially this correlation was present.  

Reconnection diffusion can explain other observations as well. For instance, 
Crutcher (2012) analyzed an extensive set of published as well as unpublished Zeeman measurements and 
showed that clouds with column densities $N_H$ less than $10^{21}$ cm$^{-2}$ are subcritical, while at higher
densities they get supercritical. He noted, that that for cold HI clouds with $N_H<10^{21}$ cm$^{-2}$ the magnetic field is 
approximately $6 \mu$G, which corresponds to the value of magnetic field strength in a much more rarefied warm neutral
media. From this he concluded that the diffuse clouds had to be formed either via the compression along magnetic 
lines or that the formation of clouds proceeds selectively only at the regions of low magnetic field strength. 
In contrast to this, on the basis of our study we claim that
reconnection diffusion presents a more appealing alternative. Indeed, as we discussed above, 
the efficient diffusion of magnetic field from the
clouds should make magnetic field strength the same in cold dense clouds and surrounding warm rarefied medium.

In addition, Crutcher (2012) notes that the kinetic and magnetic energies of such low density clouds are approximately in equipartition
and larger that the thermal energy. From the point of view of our analysis, this is a signature of the transAlfvenic supersonic turbulence indicating that the turbulence should be
efficient in driving reconnection diffusion. From the plot presented in Crutcher (2012), namely, figure 7 in his paper, it is 
evident that the
majority of the clouds in his compilation still preserves the same magnetic field strength of the order of  $6 \mu$G even
as the column density gets as high as $10^{23}$ cm$^{-2}$. We believe that this is the consequence of the
reconnection diffusion being efficient at those densities. We see a clear tendency of the increase of the mean magnetic field
in the sample for densities larger than $10^{23}$ cm$^{-2}$. A possible explanation for this is that for those clouds self-gravity gets 
important and therefore the reconnection diffusion fails to remove magnetic field from the contracting clouds fast enough. 
Note, that the high density clouds tend to be smaller and therefore, as we discussed in \S 5.2 the turbulent scales involved in the reconnection diffusion get also smaller. The reconnection diffusion for such clouds slows down.

We note that the observational results in Crutcher (2012) on the independence of magnetic field strength on the cloud
density (and the reconnection diffusion concept that accounts for these results) can be related to the empirical Larson relations (Larson 1981) obtained for interstellar turbulence. Larson (1981) found that the velocity dispersion is proportional to the square root of the cloud size, i.e. $\sigma_{V}\sim R^{1/2}$ and that the 3D density of cloud is inversely proportional to cloud size, i.e. $\rho \sim R^{-1}$. For instance, one can assume a rough equality between the kinetic energy and magnetic energy
\begin{equation}
\frac{B^2}{8\pi}\sim \rho \sigma_{v}^2,
\label{equi}
\end{equation}
which is natural for transAlfvenic turbulence as well as
 the virialization of a cloud
\begin{equation}
\frac{GM}{R}\sim \sigma_v^2   .
\label{vir}
\end{equation}
Combining Eqs.~(\ref{equi}) and (\ref{vir}) with a simplest estimate of cloud mass  $M\sim \rho R^3$ one gets
\begin{equation}
\sigma_v\sim B^{1/2} R^{1/2} 
\end{equation}
and
\begin{equation}
\rho\sim B R^{-1},
\end{equation}
which reproduce the Larson (1981) relations if the reconnection diffusion keeps magnetic field uncorrelated with density. For
cloud cores where reconnection diffusion is not fast enough to remove magnetic field on the time of the dynamic collapse (see
Tafalla et al. 1998, Reiter et al. 2011),
the Larson relations fail, in agreement with observations and simulations (see Nakamura \& Li 2011). Quantitatively, the
criterion for the reconnection diffusion to be able to remove the magnetic field from the collapsing cloud is $V_{infall}<\kappa/l_{cloud}$, where
$\kappa$ is one of the diffusion coefficients the choice of which depends on the regime of the turbulence (see \S 5.1).

\subsection{Core and envelope magnetization}

\begin{figure*}
\center
\includegraphics[width=0.9\columnwidth]{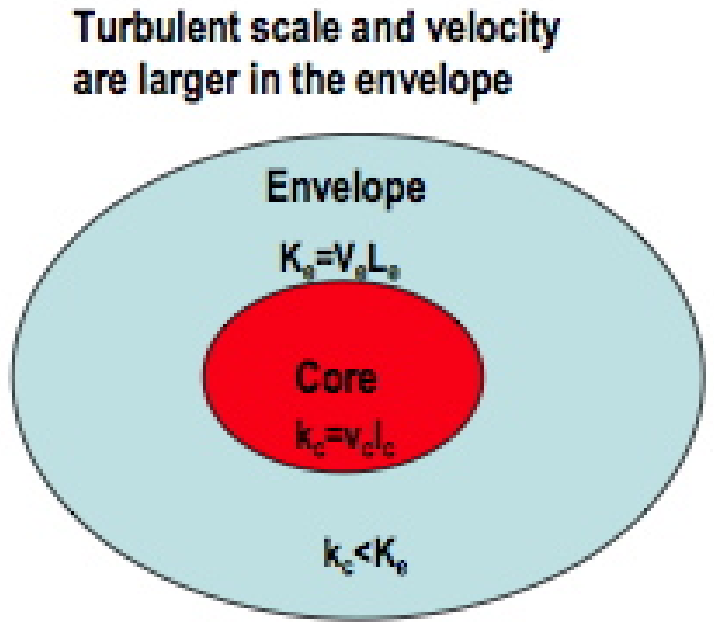}
\includegraphics[width=0.9\columnwidth]{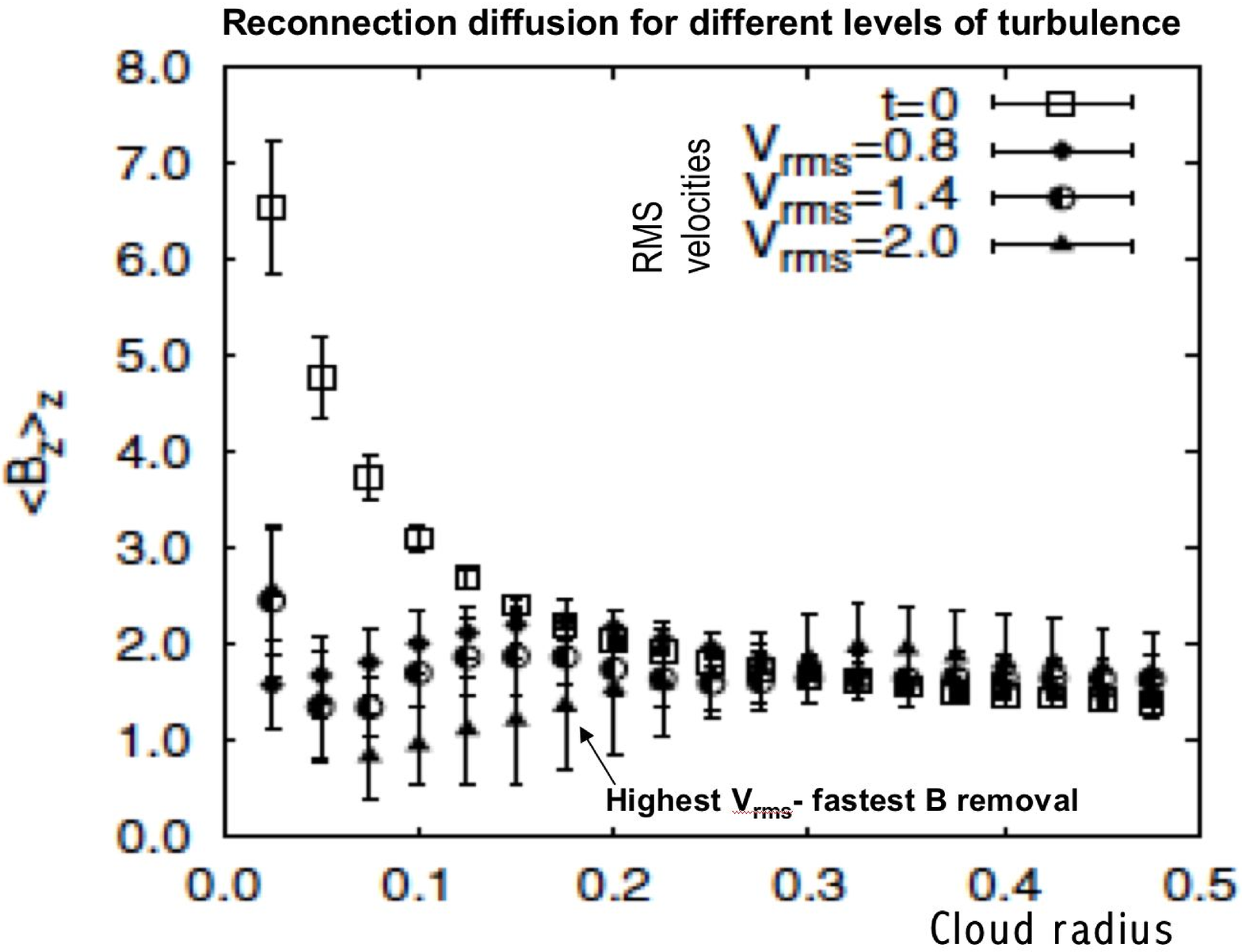}
\caption{{it Left panel}. Schematic of the cloud core and envelope in Crutcher et al. (2010). Both scale of the turbulence and velocity (and thus reconnection diffusion rate) are larger in the envelope than in the core. {\it Right panel}. Results of 
simulations (modified from Santos-Lima et al. 2010) The removal of magnetic field via reconnection diffusion increases with the increase of the turbulence level.
\label{crutcher}}
\end{figure*}

Recently the idea of star formation being mediated by ambipolar diffusion was challenged in Crutcher et al. (2010, 2011).
The authors measured matter magnetization in the core of a dark cloud and in the envelope surrounding the cloud (see our schematic representation in Figure \ref{crutcher}). 
Contrary to the predictions of the theory based on ambipolar diffusion, the observations showed that the core
is more magnetized that the envelope. Naturally, this work, which challenged the key predictions of the ambipolar diffusion
paradigm induced intensive controversy (Mouschovias \& Tassis 2009, 2010). 
Without getting into details of the well-publicized arguments of both sides, let us pose a question of
whether the results observed by Crutcher  et al. (2010, 2011) are consistent in the presence of reconnection diffusion. 

As we mentioned earlier, in the presence of gravity reconnection diffusion allows density to concentrate towards the gravitational
center, while magnetic field leave the gravitational potential. The reconnection diffusion rate and thus the rate of magnetic field removal is expected to be proportional to the level of turbulence (see Eqs. (\ref{supA}), (\ref{kappa_st}), (\ref{kappa_cl})). Simulations that illustrate this effect are
presented in Figure \ref{crutcher}. It is known that the velocities
in cloud envelopes are larger than the velocities in the cloud cores (see Taffalla et al. 1998). In addition, the scale
of turbulence involved in reconnection diffusion is also larger. Thus we expect the diffusion coefficient for the envelope (see Eq. \ref{kappa_st}) to be larger for the envelope compared to the core and faster removal of magnetic field from the envelope than from the core.  
This agrees with the results in Crutcher et al. (2010). 

\subsection{Predictions for the big picture of star formation}

Both reconnection diffusion and ambipolar diffusion remove magnetic field and can initiate gravitational collapse of the gas. However, there are substantial differences between the two processes. 

The characteristic diffusion time for magnetic fields is given by $\kappa_m/L^2$, where $\kappa_m$ is given by Eqs. (\ref{supA}), (\ref{kappa_st}), (\ref{kappa_cl}) depending on the regime of turbulence.  The diffusion coefficient   $\kappa_m$ varies spatially, as the turbulence changes from place to place and also with density. For individual clouds and cores if we neglect, for the sake of simplicity, the effects of stratification  the reconnection diffusion becomes the function of turbulent driving. For subAlfvenic driving one gets a suppression factor of $M_A^3$, which reflects the inefficiency of weak turbulence to induce magnetic field diffusion. For superAlfvenic and transAlfvenic driving, the magnetic field is to be transported with the turbulent diffusivity rate. The length scale of turbulence depends on the sources of the turbulence. If the turbulence is driven externally, the corresponding length-scale is expected to be of the order of the cloud size, if the stratification of the cloud is negligible.  

 In diffuse interstellar gas and Giant Molecular Cloud (GMC) the reconnection diffusion is expected to dominate.  One may provide evidence in support of the reconnection diffusion concept. For instance, unlike ambipolar diffusion, the reconnection does not depend on atomic level or dust physics. It was pointed to us by Bruce Elmegreen (private communication) that it is observed that the star formation is about the same in galaxies with low metallicities as in galaxies with high metallicities, which is hard to understand if magnetic field loss is governed by ambipolar diffusion. Indeed, the latter is supposed to be much faster in low-$Z$ galaxies with high metallicities and, within the ambipolar diffusion paradigm it is expected, contrary to observations, that the Initial Mass Function (IMF) and star formation rate to shifted in low-$Z$ galaxies. This is not a problem for the reconnection diffusion which is controlled only by the media turbulence. 

Within the standard star formation paradigm the low efficiency of star formation is bottlenecked by the ambipolar diffusion rate. This paradigm has been altered recently and the low efficiency of star formation was attributed to the low fraction of the cloud 
mass that has sufficiently high density to form stars (Elmegreen 2007, Vazquez-Semadeni et al. 2009). In the latter picture, star formation blows the cloud cores apart, while the efficiency of star formation is high in the cloud core with up to third of the mass going into stars. At the same time, in terms of the mass of GMCs the efficiency is low e.g. less than five percent. In other words, the cloud envelope is passive in terms of star formation, the matter is being dispersed or/and pushed aside without forming stars.  In the GMC envelopes, the reconnection diffusion is going to be efficient making sure that density and magnetic fields are well mixed. On the available
time scales it will tend to equalize the magnetic field strength within the clouds and the ambient interstellar medium, inducing the loss of part of the flux captured in the cloud at the stage of its formation. An important difference with the ambipolar diffusion is that the reconnection diffusion is not only associated with the removal of the magnetic field, but also with its turbulent mixing which tends to make the distribution of magnetic field uniform. Turbulent mixing helps keeping the diffuse media in magnetized subcritical state. In this situation, the external pressure is important for initiating collapse and which well corresponds to the observations of numerous small dark clouds not forming stars in the inter-arm regions of galaxies (Elmegreen 2011). The reconnection diffusion process is expected to relax the sharp local changes of magnetic field direction,
thus explaining the alignment of magnetic field of molecular clouds cores with the magnetic field of the spiral arms (Li \& Henning 2011).  

Ambipolar diffusion depends on the ionization of material and this results in the introduction of characteristic density for the clouds and cores to become leaky to magnetic field. In contrast, the reconnection diffusion concept implies that in realistic turbulent media there is no characteristic density for the collapse to be initiated. Therefore any cloud with the appropriate virial parameter (see McKee \& Zweibel 1992) can form stars. The difference between different clouds stems from the density controlling the
timescale of the collapse, level of steering turbulence or temperature of media. Directly, the requirement of clouds to be molecular is not present for the reconnection diffusion to induce star formation. However, molecular clouds have lower temperature. 

We would like to emphasize that the role of turbulence in the presence of reconnection diffusion is two fold. First of all, it allows the removal of magnetic flux, stimulating star formation within a contracting cloud. However, solenoidal steering also changes the virial balance making clouds less prone to collapse.
The amplitude of the steering motions increases as $v_l\sim l^{1/3}$ making larger, e.g. diffuse atomic clouds, not eligible for a gravitational collapse. The compressible motions associated with turbulence at the same time increase the density and may stimulate the collapse. The
ratio between compressible and incompressible motions depends on turbulence driving. However, simulations (e.g. Cho \& Lazarian 2002)
testify that the most of energy tend to reside in solenoidal motions. Therefore, one would expect that the increase of the level of turbulence decreases star formation, although it cannot shut it down completely. A further discussion of this issue is beyond the scope of this paper. It is clear that reconnection diffusion reveals a new important role that is played by interstellar turbulence.

\subsection{Reconnection diffusion and extreme cases of star formation}

Reconnection diffusion concept provides new ways of approaching challenging physics of star formation
in extreme environments.  For instance, galaxies emitting more than $10^{12}$ solar luminosities in the far-infrared are called ultra-luminous infrared galaxies or ULIRGs. The physical conditions in such galaxies are extreme with very high density of 
cosmic rays (see Papadopoulos et al. 2011). Ambipolar diffusion is expected to be suppressed due to cosmic ray
ionization. At the same time these environments provide the highest star formation rate which is suggestive of a process
which removes magnetic fields irrespectively of the level of ionization. Reconnection diffusion is such a process. 

The formation of early stars  (see Chiappini et al. 2011) is a great problem for which the effects of magnetic fields
are debated. Reconnection diffusion mitigates the effects of magnetic fields and therefore decreases the uncertainties
associated with the presence of magnetic fields at the sites of early Universe star formation. 

\section{Additional consequences of reconnection diffusion}

\subsection{Reconnection diffusion and cosmic ray acceleration in dark clouds}

\begin{figure}
\center
\includegraphics[width=0.45\textwidth]{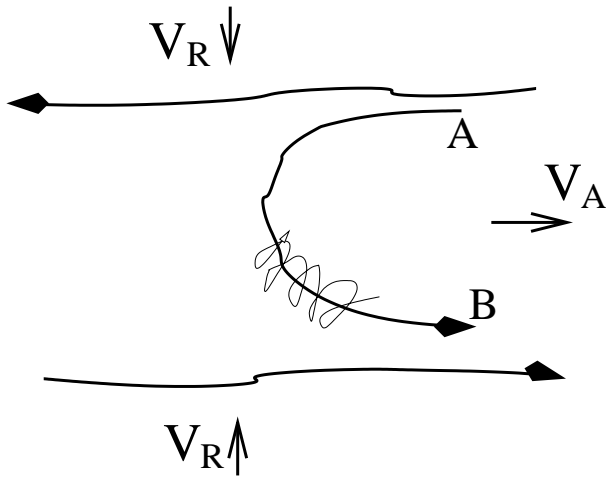}
\caption{First order Fermi acceleration as cosmic rays bounce within a 3D loop of reconnected flux that shrinks due to magnetic reconnection.
From Lazarian 2005.
\label{CR}}
\end{figure}

The lower energy e.g. $\sim 100$ MeV cosmic rays dominate the ionization of cool neutral gas, especially dark
UV shielded molecular regions (Goldsmith \& Langer 1978). Some observations indicate substantial changes
of the cosmic rate ionization rate between diffuse and dense gas (see McCall et al. 2003) which may
be the consequence of local cosmic ray acceleration. As we discussed, reconnection diffusion invokes LV99 model of reconnection.
However, the same type of reconnection is expected to
accelerate cosmic rays (de Gouveia dal Pino \& Lazarian 2005, Lazarian 2005)\footnote{The predicted spectrum 
without taking the backreaction of the accelerated particles is
$N(E)dE\sim E^{-5/2} dE$. Considerations in Drake et al. (2006) suggest that the spectrum of the particles
can get shallower if the backreaction is taken into account.}. The acceleration happens
as particles bounce back and forth within shrinking magnetic loops (see Figure \ref{CR}). Recently, the acceleration of
cosmic rays in reconnection has been invoked to explain results on the anomalous cosmic rays obtained by
Voyager spacecrats (Lazarian \& Opher 2009, Drake et al. 2010), the local anisotropy of cosmic rays (Lazarian \& Desiati
2010) and the acceleration of cosmic rays in clusters of galaxies (Lazarian \& Brunetti 2011). Naturally, the
process of acceleration is much wider spread and not limited by the explored examples.

The maximal energy of the particles can be estimated with the standard formulae (see Longair 2011)
$E_{max}\sim 10^{16} eV \hat{B} \hat{R}$, where magnetic field $\hat{B}$ is normalized by  1 $\mu$G
and the radus is normalized by 1 $pc$. This maximal energy may not be reached due to the CR losses
if the acceleration is insufficiently efficient. Numerical simulations by Kowal et al. (2012) show a
relative inefficiency of first order Fermi acceleration by reconnection in pure turbulence without a large scale
magnetic reversal. This is likely to be due to the gentle variations of magnetic field in the GS95 model of
turbulence. As we discussed earlier (see \S 5.1) reconnection in strong MHD turbulence does necessarily 
take place, but the reconnecting bundles of magnetic field cross each other at a small angle $\phi$ and the
resulting rate of energy gain $V_A\cos\phi/c$ is low. Gravitational forces acting on a 
collapsing core sufficiently perturb magnetic field
lines increasing the efficiency of the acceleration within reconnection regions.

Stretching of the magnetic field lines happens both inside molecular clouds and outside them as curved magnetic
field lines diffuse away. The Alfven velocity outside clouds is larger and this may provide a more efficient
acceleration in agreement with the ionization results for the $\zeta$ Persei cloud (see Le Petit 2004). Naturally,
more 3D modeling\footnote{Modeling in Kowal et al. (2011b) showed that the acceleration in 2D and 3D proceed
at a different rate, which questions the applicability of results obtained in 2D simulations (see Drake et al. 2010) to the actual astrophysical systems.} of the acceleration in reconnection regions is necessary to seek the quantitative agreement
with observations.

\subsection{Heating of clouds} 

Ambipolar drag induces additional heating of star forming clouds, which can potentially be detected as the excess of 
the heat after turbulent, cosmic ray etc. ways of heating are accounted for. As any fast reconnection model, LV99 predicts that only
a small fraction of magnetic energy is going to be dissipated directly through Ohmic heating. Most of the energy is going to be 
consumed by the turbulent outflow. Thus in the process of reconnection diffusion we expect that the heating is going to
be linked with turbulence in clouds. Additional turbulent energy input might be detected through the variations of
the power spectrum. The latter can be measured by the Velocity Channel Analysis (VCA) or Velocity Correlation Spectrum
(VCS) techniques (Lazarian \& Pogosyan 2000, 2004, 2006, Padoan et al. 2004, 2009, Chepurnov \& Lazarian 2009, Chepurnov et al. 2010, see
Lazarian 2010 for a review).

\subsection{Star formation simulations: criterion for representing magnetic diffusion}

We believe that flux freezing being the dominant idea in astrophysical community strongly influenced the interpretation of the results of numerical simulations. Indeed, within the flux freezing paradigm the only way to explain how
conductive matter become concentrated without dragging magnetic field lines along  
is to postulate its motion along magnetic field lines. Any other explanation would open a a pandora box of the necessity
to  justify how diffusive numerical simulations can represent highly conductive
astrophysical environments. 
 
The concept of reconnection diffusion provides a new outlook at the existing simulations. Plasmas and magnetic 
field are subject to efficient reconnection diffusion both in astrophysical conditions and in numerical simulations. As
we discussed in \S 4.2  and \S 6.4 the rate of reconnection diffusion is independent of the detailed microphysics
of local reconnection events. In this respect, numerical small scale reconnection does not compromise the results
of numerical simulations of turbulent environments. However, at scales where numerical diffusion suppresses 
turbulence the physics reproduced by codes is different in simulations and in real astrophysical conditions.

The field wandering weakly which is essential for the LV99 reconnection depends on the amplitude and scale of
turbulent motions and does not depend on the small scale microphysics. These is excellent news for
MHD simulations of astrophysical turbulent environments, in particularly, the star formation. As
we mentioned in \S 8.1 astrophysical simulations
are not in the right ballpark of the magnetic Reynolds or Lundquist numbers and it is important that in terms of
magnetic reconnection this huge difference do not alter the physics of the process. 

In fact, reconnection diffusion has always been part of the numerical simulations involving turbulence even if its role 
has not been identified. For instance, Julian Krolik pointed to us his work on magnetic field transport around a black hole (Beckwith, Hawley \& Krolik 2009). We also believe that reconnection diffusion is the
reason for the expulsion of the magnetic flux in the simulations of Romanova et al. (2011). We might speculate the simulations
in Li et al. (2012) are also affected by reconnection diffusion. We may argue that some effects  explained as
the consequence of the accumulation of matter along magnetic field lines (see \S 11.6) are also due to reconnection diffusion.
However, without detailed studies of the aforementioned simulations it is not clear whether the real reconnection diffusion physics or bogus effects of numerical diffusivity are at play. For instance, Crutcher et al. (2011) refers to the simulations in Luntilla et al. (2009) which produce, in 
agreement with observations, higher magnetization of the cloud cores. If these cores are of the size of several grid units across, numerical
effects rather than reconnection diffusion may be dominant and turbulence is suppressed at these scales. 

All in all, one can formulate the criterion for the numerical simulations to represent the actual diffusion of magnetic fields: {\it if
on the scales of study simulations exhibit turbulence, it is reconnection diffusion that dominates the numerical one}. If this criterion is
satisfied, in terms of the magnetic field diffusion, the simulations are trustworthy. Thus we claim that if the cores in the simulations
are turbulent, the removal of magnetic flux from them via reconnection diffusion is similar to astrophysical highly conductive environment.
This opens a possibility of developing adaptive mesh codes designed to reproduce astrophysical magnetic diffusion correctly in spite
of the difference in Lundquist numbers.

\section{Discussion}

\subsection{Physics of reconnection diffusion}

The idea that fast reconnection predicted by LV99 model can change the picture of star formation by 
inducing magnetic field diffusion has been discussed in a number
of papers starting with Lazarian (2005). However, the physical picture of the process has not been
properly discussed. This was a problem, as the competing picture of magnetic drift via ambipolar
diffusion is very clear in terms of the physics involved (see Mouschovias 1991). 

The above shortcoming is dealt with in the present paper where two physical pictures of the process, 
macrophysical based on LV99, and microphysical, based on ELV11, are presented. The two approachers
are equivalent in the very core of the underlying physics, but in terms of understanding of the process
they exhibit complementary features. Indeed,  LV99 primary deals with large scale fluxes that
intersect and reconnect. This is the standard way of dealing with magnetic reconnection. The large scale
variations of magnetic field must be dealt in star formation and the answer provided by LV99 is that such
variations do not stop the process of reconnection diffusion. Naturally, as we also showed in this paper,
the LV99 model can be applied to different scales of magnetic field hierarchy, including the hierarchy of the
turbulent eddies within the magnetized flow. At this point it enters the domain of self consistently describing
the cascade of reconnection events in the turbulent flow that is important for the continues changes of
magnetic field line identities that is essential for reconnection diffusion. This is also the domain which
is described by the concept of "spontaneous stochasticity" (see ELV11). The latter, instead of describing the
constantly and stochastically changing magnetic field line connections, uses the description which 
explicitly appeals to the magnetic field stochasticity at turbulent scales and the violation of flux freezing. 

The gain in physical understanding of the reconnection diffusion enabled us to provide a better
description of the consequences of the process of star formation. For instance, it allowed us to 
better identify the role of weak and strong turbulence
for transporting magnetic fields. The estimates that we provided can be tested both observationally and
numerically. In view of the latter point, we would like to stress that by itself the existing numerical
studies of reconnection diffusion cannot be used to prove the concept. The simulations so far have been
done in MHD regime and instead of the generalized Ohm's law (see Eq. (\ref{Ohm}) have used ordinary
resistivity. In this situation it is the LV99 reconnection theory and the
 physically equivalent to it theory of spontaneous stochasticity (ELV11) that justify that why
one can disregards the small scale physics dealing with the
 reconnection diffusion process.    
  
 Our results in the paper should not be understood as the claim that ambipolar diffusion is unimportant
 for star formation. Future research, both observational and theoretical, should determine the relative 
 importance of reconnection and  ambipolar diffusion processes. 
 As observations show that turbulence is ubiquitous in diffuse ISM and molecular clouds, we expect the 
 reconnection diffusion to be important for magnetic field transport in these environments. However, in very
 quiescent cloud cores ambipolar diffusion may be the dominant process.  
 
 \subsection{Intuitive understanding of reconnection diffusion}

The idea of magnetic flux freezing is so deeply rooted in astrophysics, that any attempts to challenge it 
sound rather heretical. We would like to stress that fast reconnection in turbulent magnetized media is very natural. Without it
the turbulent fluid would create felt-like structures, as was advocated by Don Cox (private communication). 
Indeed, intersecting and not being able to pass through each other magnetic field lines are bound to 
stop magnetized fluid behave like a fluid. If, however, this is not the case and magnetized fluids preserve fluid-type
behavior in the presence of turbulence, one has to accept the efficient
diffusion of matter and magnetic fields. This happens through reconnection diffusion as we explained in this paper.
It worth mentioning that the ideas of magnetic field meandering that have been invoked for decades
to understand the observed diffusion of cosmic rays perpendicular to magnetic field (see Parker 1965,
Jokipii 1973) are naturally related to the reconnection diffusion concept.

As we discussed in the paper, the reconnection diffusion is closely related to the flux freezing violation in turbulent 
environments. 
The failure of flux freezing in turbulent fluids has very big astrophysical consequences for star formation and
beyond it. The widely accepted point of view is that in astrophysical situations the flux freezing is "nearly" fulfilled and
the violations are due to the existing finite non-ideal effects. The problem is that this is not true in realistically turbulent astrophysical fluids. However, one should keep in mind that turbulence, unlike resistivity, does not destroy magnetic field lines, but it makes the magnetic field stochastic.
The total magnetic flux does not change, but the charged particles get the possibility of exploring the entire volume, which
also means that magnetic field gets diffusive. In turn, the latter entails magnetic field filling the entire volume
and mitigating its effects of counteracting gravitational compressing the matter. 

\subsection{Turbulent and magnetic support of clouds}

The traditional approach to star formation frequently appeals to turbulent and magnetic support of the clouds (see Mestel 1965).
As magnetic field is usually assumed to be in rough equipartition with kinetic motions, one may wonder whether
neglecting magnetic support just amounts to the factor of order unity in the picture above. 

There is, however, a serious difference between the effects of magnetic field and turbulence. For the 
virial support by turbulence, only motions less than the cloud scale are important. The larger scale motions do not
enter the virial equation. The outside cascade, if anything, can compress the cloud due to compressible
turbulent fluctuations and cloud-turbulent interactions. The role of magnetic field is different. If the large scale magnetic
field is dragged into the cloud, it can only be amplified by the cloud compression. Therefore the large
scale equipartition between turbulence and magnetic field does not preclude the magnetic field to be dominant on the scale
of self-gravitating cores.

The process of reconnection diffusion allows the magnetic field to equalize inside and outside the cloud, decreasing
the effect of magnetic support. This corresponds to the modern understanding of star formation as a very dynamic process
with no necessity of support of small scale infall of matter (see Elmegreen 2011 and references therein). 

Incidentally, reconnection diffusion shows
that the textbook picture of the magnetized cloud with magnetic field support in the direction perpendicular
to magnetic field and turbulence providing the vertical extend of the cloud is not sustainable. Reconnection
diffusion is expected to remove the excess of magnetic field from the cloud on the dynamical time scales. 
The conventional picture above holds only in the absence of strong MHD turbulence.  

\subsection{Reconnection diffusion and turbulent ambipolar diffusion}

An interesting study focused on the ambipolar diffusion physics in a turbulent flow was performed Heitsch et al. (2004, henceforth HX04).
They performed 2.5D simulations of turbulence with two-fluid code and examined the decorrelation of neutrals and magnetic field
that was taking place as they were driving the turbulence. The study reported an enhancement of ambipolar diffusion rate compared
to the ambipolar diffusion acting in a laminar fluid. HX04 correctly associated the enhancement with turbulence
creating density gradients that are being dissolved by ambipolar diffusion (see also Zweibel 2002). Due to magnetic field being perpendicular to the flow in 2.5D, the set up precluded magnetic field reconnection and magnetic fields preserved their identity 
throughout the simulations (cf. \S 6). Thus HX04 studied the effect different from reconnection diffusion that we deal in
the paper. They termed the process "turbulent ambipolar diffusion".

The set up in HX04 presents a special case of a magnetized flow when magnetic fields act only as an additional
pressure within the fluid that exhibits hydrodynamic behavior irrespectively of the strength of magnetic field. Therefore, similar to the hydrodynamic case, turbulence in this situation cannot be weak, even if $V_A\gg V_L$ (cf. Table 1) and 2.5D eddies stir the fluid irrespectively of the strength of the driving. Ions stayed entrained on the field lines (cf. Figure \ref{regimes})
while both neutral density and magnetic field spread diffusively at approximately the eddy rate. The decorrelation arose due to small scale ion-neutral drifts. 

Both viscosity and the decorrelation of magnetic field and neutral density in HX04 are due to the slippage of neutrals
and ions and therefore in the presence of hydrodynamic-type eddies the diffusivity given by Eq. (\ref{hydro}) agrees well with their finding.
In analogy with hydrodynamics, we expect that the diffusivity would not change even as ambipolar diffusion rate changes. Indeed, 
if the rate of ambipolar diffusion gets smaller the turbulent cascade proceeds to smaller scales allowing mixing at those scales\footnote{A similar process takes place in the case of molecular diffusivity in turbulent hydrodynamic flows. The result for the latter flows is well known: in turbulent regime molecular diffusivity is irrelevant
for the turbulent transport. Indeed, in the case of high microscopic diffusivity, the turbulence provides mixing down to a scale $l_1$ at which the microscopic diffusivity both, suppresses the cascade and ensures efficient diffusivity of the contaminant. In the case of low microscopic diffusivity, turbulent mixing happens down to a scale $l_2\ll l_1$, which ensures that even low microscopic diffusivity is sufficient to provide efficient diffusion. In both cases the total effective diffusivity of the contaminant is given by the product of the turbulent injection scale and the turbulent velocity.}. The
reduced ambipolar diffusion would still be adequate to decorrelate the magnetic field and the reduced scales. We infer that the limiting case of this flow is the 2.5D flow with no ambipolar diffusion but with the diffusivity still given by Eq. (\ref{hydro}). This still would not be the reconnection diffusion case, as no reconnection is allowed.   
  
The difference of what we discussed in the present paper and the idea in HX04 arises from the difference in the flows that we considered.
The set up that we deal with is a generic 3D turbulent flow where magnetic reconnection is inevitable. Therefore, as we discussed in the
paper, ions themselves are diffusive (see \S 6) and the {\it ionic density decorrelates with magnetic field} as well. 
Similarly to what we discussed earlier, in the absence of ambipolar diffusion, the turbulence propagates to smaller scales making small-scale interactions possible. On the other hand, ambipolar diffusion affects the turbulence, increasing its damping. As a result, analogously to 2.5D case above, the ambipolar diffusion acts in two ways, in one to increase the small-scale diffusivity of the magnetic field, in another is to decrease the turbulent small-scale diffusivity and these effects essentially compensate each other\footnote{A possible point of confusion is related to the difference of the physical scales involved. If one associates the scale of the reconnection with the thickness of the Sweet--Parker layer, then, indeed,
the ambipolar diffusion scale is much larger and therefore the reconnection scale gets irrelevant. However, within the LV99 model of reconnection, the scale of reconnection is associated with the scale of magnetic field wandering. The corresponding scale depends on the turbulent velocity and is not small.}. Therefore, we believe that, at least for the case of strong MHD turbulence, ambipolar diffusion does not play any role for the turbulent
transport in magnetized fluid. In a generic situation of 3D magnetized turbulence, reconnection is essential and reconnection diffusion takes place. The diffusion coefficient $\sim LV_L$ corresponds to our prediction of reconnection diffusion
induced by transAlfvenic and superAlfvenic turbulence. We note that, in the presence of turbulence, the independence of the gravitational collapse from ambipolar diffusion rate was reported in numerical simulations by Balsara, Crutcher \& Pouquet (2001). 
 
 What does happen in the subAlfvenic case of weak turbulence? As we mentioned before, we are not aware of the studies
of diffusion in two fluids in this regime. At the same time,
 this may be potentially the most interesting case as far as the interplay of turbulence and ambipolar diffusion is concerned.
 As we discussed in \S 3.2 and \S 5.1 MHD turbulence at large scales corresponds to the "weak" regime and
can be viewed as a collection of non-linear weakly interacting waves. However, at a smaller scale, namely at the scale
$l_{trans}$ given by Eq. (\ref{trans}) the turbulence gets into the regime of strong interactions, when the intensive mixing happens in the 
direction perpendicular to the local direction of magnetic field. The  diffusivity associated with turbulence is given by Eq. (\ref{kappa_st}). If the ambipolar diffusivity is less than this value, it will not play any role and the diffusivity will be purely ``turbulent''. If, however, 
damping happens at scale larger than $t_{trans}$ a hypothetical new regime of "weak reconnection diffusion" may be present (see \S 7).
The study of "turbulent ambipolar diffusion" with weak MHD turbulence
has not been performed as far as we know. The effects of the enhancement of the total diffusivity are thus unclear.
We might expect little, if any, parameter space for the ``turbulent ambipolar diffusion'' when turbulence and ambipolar diffusion synergetically enhance diffusivity, acting in unison. Nevertheless, this point should be tested by three-dimensional two-fluid simulations exhibiting both ambipolar diffusion and turbulence. The effect to search in order to test the effect of "turbulent ambipolar diffusion" is the enhancement of the diffusivity compared to the rate of diffusivity arising from the weak subAlfvenic turbulence (given by Eq. (\ref{kappa2})).

\subsection{Reconnection diffusion and hyper-resistivity}

To explain fast removal of magnetic field from accretion disks Shu et al. (2006) appealed to the hyperrestivity concept (Strauss
1986, Bhattacharjee \& Hameiri 1986, Hameiri \& Bhattacharjee, Diamond \& Malkov 2003). The studies introducing
hyperresistivity attempt
to derive the effective resistivity of the turbulent media in the context of the mean-field resistive MHD. Using magnetic helicity
conservation the authors derived the electric field. Then, integrating by part, one obtained a term which could be identified with 
effective resistivity proportional to the magnetic helicity current. There are several problems with this derivation. In particular,
the most serious is the assumption that the helicity of magnetic field and the small scale turbulent fields are separately conserved,
which erroneously disregard the magnetic helicity fluxes through open boundaries, which is essential for fast stationary reconnection (see more discussion in Kowal et al. 2009 and ELV11). 
In more general terms, hyper-resistivity idea is an incarnation the mean-field approach to explaining fast reconnection. As explained in ELV11, the problem of such approaches is that the lines of the actual astrophysical magnetic field should reconnect, not the lines of the mean field. Therefore the correct approach to fast reconnection should be independent of the spatial and
time averaging.
   
All in all, we believe that the concept of hyperresistivity is poorly justified and should not be applied to astrophysical environments.

\subsection{Reconnection diffusion and collecting matter along magnetic field}

As we mentioned in \S 9.1 an alternative way for changing flux to mass ratio is to allow conducting matter to be accumulated along
magnetic field lines. This process definitely takes place, but the prescription of one dimensional motion of matter is very
restrictive. More importantly, as we discussed in \S 6 the idea of fixed magnetic field lines is not applicable to turbulent magnetized fluid.
Thus the effects of reconnection diffusion should inevitably interfere even if plasmas is launched along magnetic field lines.  

For instance, our work on reconnection diffusion in diffuse interstellar medium should  also be distinguished from the research on the de-correlation of magnetic field and density within compressible turbulent fluctuations. Cho \& Lazarian (2002, 2003) performed three-dimensional MHD simulations and reported the existence of separate turbulent cascades of Alfven and fast modes in strongly driven turbulence as well as a cascade of slow modes driven by Alfv\'enic cascade. Slow modes in magnetically dominated plasma are associated with density perturbations with marginal perturbation of magnetic fields, while the same is true for fast modes in weakly magnetized or high beta plasmas. Naturally, these two modes de-correlate magnetic fields and density on the crossing time of the wave. This was the effect studied in more detail in one-dimensional setting both both analytically and numerically by Passot \& Vazquez-Semadeni (2003), who stressed that the enhancements of magnetic field strength and density may correlate and anti-correlate in turbulent interstellar gas within the fluctuations and this can introduce the dispersion of the mass-to-flux ratios within the turbulent volume. Each of the fluctuations provide a {\it transient} change of the pointwise magnetization. In the absence of other effects, e.g. related to the thermal instability, the de-correlation is reversible. In comparison, the ``reconnection diffusion'' deals with the {\it permanent} de-correlation of magnetic field and density making magnetic field-density de-correlation irreversible. In many instances both processes act together providing the observed (see \S 9.1)  decorrelated magnetic field and density state. 

While for the diffuse media both collecting the matter through compressible modes and mixing through reconnection diffusion can act together,
the idea of star formation based on collecting matter along magnetic field lines is more problematic.  
The scales for such one-dimensional collection are enormous ($\sim 1$ kpc see Vazquez-Semadeni et al. 2011) and it is not feasible that in
turbulent environments reconnection diffusion would not
interfere with the postulated one-dimensional motion.  We believe that
reconnection diffusion is an intrinsic part of the simulations of the turbulent interstellar medium and it should be accounted for
in interpreting the results of numerical simulations provided that the criterion in \S 10.3 is satisfied. If the criterion is not
satisfied, then the magnetic diffusivity is dominated by numerical effects and one should be cautious interpreting such simulations.

\subsection{Reconnection diffusion and modern understanding of MHD turbulence and reconnection}

The concept of reconnection diffusion is deeply rooted in the modern understanding of MHD turbulence and its
intrinsic connection with reconnection. While mixing motions perpendicular to the local direction of magnetic field
is a part of the GS95 picture of Alfvenic turbulence, the star formation textbooks frequently depict magnetic turbulence as a collection of waves with wavevectors mostly parallel to magnetic field. The latter induce only marginal mixing of matter
and therefore other mechanisms of changing flux to mass ratio are required.

Mixing motions inevitably induce the question of magnetic reconnection. Fortunately, as we discussed at the end of
section \S 4.3 the LV99 model provides the necessary rates thus providing
a physical justification for both the GS95 model and the reconnection diffusion concept.

In the paper we considered MHD turbulence where the flows of energy 
in the opposite directions are balanced. When this is not true, i.e. when the turbulence has non-zero 
cross-helicity, or imbalanced (see \S 4.3). Studies of
reconnection diffusion in such turbulence is a goal for a future.
In terms of star formation, we do not believe that turbulence is strongly imbalanced. In fact,
in compressible media 
the imbalance decreases due to reflecting of waves from pre-existing density fluctuations and 
due to the development of parametric instabililites (see Del Zanna, Velli \& Londrillo 2001).

\conclusions

In this paper we consider the process of reconnection diffusion which arises in magnetized turbulence
due to fast magnetic reconnection. We consider GS95 model of strong turbulence and
LV99 model of fast magnetic reconnection. We claim that in the presence
of strong magnetic turbulence the notion of plasma frozen onto
a magnetic flux tube becomes meaningless. Due to reconnection, magnetic field lines constantly change their identity inducing intensive mixing of plasmas. The intensity of this process is determined by the intensity of turbulence
and does not depend on microphysics of reconnection, e.g. it does not depend on the collisional or collisionless nature
of the small scale reconnection events. 

While reconnection diffusion is important for many key astrophysical processes, e.g. heat transport in plasmas, generation of magnetic field etc., this paper is focused on star formation, where the flux freezing idea is at the core of the existing paradigm. Therefore the concept of reconnection
diffusion alters the paradigm. The new concept explains the existing observational data that contradicts the theory based on
ambipolar diffusion. In particular it can explain data on the decorrelation of density and magnetic field in diffuse
interstellar media, fast removal of magnetic field in molecular clouds, higher magnetization of cloud core compared to envelope, properties of circumstellar rotation disks, independence of star formation rate on metallicity, possibility of star formation
in ULIRGs, selected empirical relations etc. It also provide predictions for future observations and helps to bridge the gap between numerical simulations and actual star formation in galaxies and early universe. More studies of the new approach to magnetic field diffusion in star
formation is necessary.

%
%
%
%

\begin{acknowledgements}
The research  is supported by the Center for Magnetic
Self-Organization in Laboratory and Astrophysical Plasmas. Stimulating environment provided by  Humboldt Award 
at the Universities of Cologne and Bochum, as well as a Fellowship at the International 
Institute of Physics (Brazil) is acknowledged.  A productive exchange on star formation with Bruce Elmegreen was
particularly valuable. I am grateful to  Elisabeth Gouveia dal Pino, Reinaldo Santos-Lima, Dick Crutcher,
Chris Mckee, Greg Eyink, Ethan Vishniac for stimulating discussions on various aspects of the problem. Exchanges with Julian Krolik on the reconnection diffusion around black holes and with Ellen Zweibel on the role of ambipolar diffusion are acknowledged. We thank the anonymous referee for useful input and Blakesley Burkhart for reading
the manuscript. 

\end{acknowledgements}

\end{document}